\documentclass[twocolumn,superscriptaddress,prc,nofootinbib]{revtex4-1}

\usepackage{graphicx}
\usepackage{color}
\usepackage{amssymb}
\usepackage{amsthm}
\usepackage{textcomp}
\newcommand{\dl}{\mbox{DAMA/LIBRA}}
\newcommand{\naitl}{\mbox{NaI(Tl)}}

\newcommand{\iseven}{\mbox{$^{127}$I}}
\newcommand{\keVr}{\mbox{keV$_{\text{nr}}$}}
\newcommand{\keVee}{\mbox{keV$_{\text{ee}}$}}
\newcommand{\kevee}{keV$_{\text{ee}}$}
\newcommand{\us}{$\mu$s}
\newcommand{\lipn}{\mbox{$^{7}$Li(p,n)$^{7}$Be}}

\begin{document}


\title{Scintillation efficiency measurement of Na recoils in \naitl\ below the \dl\ energy threshold}

\author{Jingke Xu}
\email[Corresponding author, ] {jingkexu@princeton.edu}
\affiliation{Department of Physics, Princeton University, Princeton, NJ 08544, USA}
\author{Emily Shields}
\affiliation{Department of Physics, Princeton University, Princeton, NJ 08544, USA}
\author{Frank Calaprice}
\affiliation{Department of Physics, Princeton University, Princeton, NJ 08544, USA}
\author{Shawn Westerdale}
\affiliation{Department of Physics, Princeton University, Princeton, NJ 08544, USA}
\author{Francis Froborg}
\affiliation{Department of Physics, Princeton University, Princeton, NJ 08544, USA}
\author{Burkhant Suerfu}
\affiliation{Department of Physics, Princeton University, Princeton, NJ 08544, USA}
\author{Thomas Alexander}
\affiliation{Amherst Center for Fundamental Interactions and Physics Department, Amherst, MA 01003, USA}
\affiliation{Fermi National Accelerator Laboratory, Batavia, Illinois 60510, USA}
\author{Ani Aprahamian}
\affiliation{Department of Physics, University of Notre Dame, Notre Dame, IN 46556, USA}
\author{Henning O. Back}
\affiliation{Department of Physics, Princeton University, Princeton, NJ 08544, USA}
\author{Clark Casarella}
\affiliation{Department of Physics, University of Notre Dame, Notre Dame, IN 46556, USA}
\author{Xiao Fang}
\affiliation{Department of Physics, University of Notre Dame, Notre Dame, IN 46556, USA}
\author{Yogesh K. Gupta}
\altaffiliation{Nuclear Physics Division, BARC, Mumbai-400085, India}
\affiliation{Department of Physics, University of Notre Dame, Notre Dame, IN 46556, USA}
\author{Aldo Ianni}
\affiliation{INFN Laboratori Nazionali del Gran Sasso, SS 17 bis Km 18\_910, 067010 Assergi (AQ), Italy}
\author{Edward Lamere }
\affiliation{Department of Physics, University of Notre Dame, Notre Dame, IN 46556, USA}
\author{W. Hugh Lippincott}
\affiliation{Fermi National Accelerator Laboratory, Batavia, Illinois 60510, USA}
\author{Qian Liu}
\affiliation{Department of Physics, University of Notre Dame, Notre Dame, IN 46556, USA}
\author{Stephanie Lyons}
\affiliation{Department of Physics, University of Notre Dame, Notre Dame, IN 46556, USA}
\author{Kevin Siegl}
\affiliation{Department of Physics, University of Notre Dame, Notre Dame, IN 46556, USA}
\author{Mallory Smith}
\affiliation{Department of Physics, University of Notre Dame, Notre Dame, IN 46556, USA}
\author{Wanpeng Tan}
\affiliation{Department of Physics, University of Notre Dame, Notre Dame, IN 46556, USA}
\author{Bryant Vande Kolk}
\affiliation{Department of Physics, University of Notre Dame, Notre Dame, IN 46556, USA}

\date{\today}

\begin{abstract}

The dark matter interpretation of the DAMA modulation signal depends on the \naitl\ scintillation efficiency of nuclear recoils. Previous measurements for Na recoils have large discrepancies, especially in the \dl\ modulation energy region. We report a quenching effect measurement of Na recoils in \naitl\  from 3\,\keVr\ to 52\,\keVr, covering the whole \dl\ energy region for light WIMP interpretations. By using a low-energy, pulsed neutron beam, a double time-of-flight technique, and pulse-shape discrimination methods, we obtained the most accurate measurement of this kind for \naitl\ to date. The results differ significantly from the DAMA reported values at low energies, but fall between the other previous measurements. We present the implications of the new quenching results for the dark matter interpretation of the DAMA modulation signal.

\end{abstract}

\keywords{DAMA-LIBRA, dark matter, \naitl, quenching effect, \naitl\ quenching factors}

\maketitle




\section{Introduction}
\label{sec:introduction}

For over a decade, the DAMA experiments (DAMA-NaI and \dl) have been observing an annual modulation in the rate of events in the low-energy region of NaI(Tl) detectors~\cite{DAMA2013_Phase1}. This modulation signal has an extremely high statistical significance (9.3$\sigma$) and is often interpreted as evidence for WIMP dark matter interactions, such as low mass WIMP scattering~\cite{Hooper2010_LightWIMP} or inelastic WIMP scattering~\cite{Smith2001_InelasticDM, Bernabei2002_InelasticDM}. Several experiments have ruled out the DAMA dark matter claim in the standard WIMP picture~\cite{XENON2011_LightWIMP, XENON2012_225days, CDMS2013_SiDM, LUX2014_DMResult}, while alternative WIMP theories might still reconcile the experimental results~\cite{Feng2011_IVDM, Chiara2015_pseudoscalar}.

The dark matter interpretation of the DAMA modulation signal depends on the scintillation efficiency of \naitl\  for sodium and iodine recoils relative to that of gammas (electron recoils); the former could be induced by WIMP scattering interactions, while the latter are used to calibrate the detectors. Nuclear recoils in \naitl, due to the small fraction of energy transfer to electrons, typically produce less scintillation light compared with electron recoils with the same energy deposition. The relative ratio is usually referred to as the nuclear recoil quenching factor. This factor is critical to translate the observed electron equivalent energy into nuclear recoil energy for dark matter analysis.

DAMA reports a quenching factor of 0.3 for sodium recoils and 0.09 for iodine recoils~\cite{DamaQuench1996}. These results were obtained by exposing a \naitl\ detector to neutrons from a $^{252}$Cf source.  The nuclear recoil signals produced by the neutrons were compared to Monte Carlo-simulated sodium and iodine recoil spectra to extract the quenching factors, which were assumed to be  energy-independent. Since then, several more experiments have been carried out to measure the sodium and iodine quenching factors as a function of nuclear recoil energy using mono-energetic neutron sources, mostly deuterium-deuterium neutron generators~\cite{Spooner1994, Tovey1998, Gerbier1999, Jagemann2006, Simon2003}. By looking for coincidence signals between a \naitl\ detector and neutron detectors at fixed neutron scattering angles, this technique could provide a direct measurement of the DAMA modulation energy scale if the signal is interpreted as nuclear recoils. A few of these measurements reported quenching factors consistent with the DAMA results, especially in energy regions higher than 20\,keV nuclear recoil energy (20\,\keVr). However, recent measurements by Chagani~\cite{Chagani2008} and Collar~\cite{Collar2013} led to new Na recoil quenching factors significantly deviating from the DAMA values over a wide energy range, and these two new measurements also conflict seriously with each other at low energies, where the DAMA modulation signals occur in a light WIMP interpretation.

In addition to these inconsistencies, the previous Na recoil quenching measurements in \naitl\ typically carry large uncertainties, ranging from $\sim$\,10\% (relative) to, at the lowest energies, over 100\% (relative). Therefore, it is necessary to conduct new quenching measurements that can significantly improve the quenching-factor accuracy and provide a reliable Na recoil calibration for the light WIMP interpretation of the DAMA results, as well as for other \naitl\ dark matter experiments~\cite{Kim2015_NaI, DMIce2014_firstdata, ANAIS2012_ANAIS0, NaIAD2005_WIMP, SABRE2013_TAUP}.

In this paper, we report a \naitl\ quenching measurement using a pulsed neutron beam produced by the FN tandem facility at the University of Notre Dame Nuclear Science Laboratory. This measurement was designed to achieve an overall uncertainty of $\sim$\,5\% and an energy threshold of a few \keVr. Several techniques were combined to suppress backgrounds and uncertainties, as summarized below:
\begin{enumerate}
\item A triple time-coincidence between a pulsed neutron beam, a \naitl\ detector and an angular array of neutron detectors was used to select neutron events and to reduce random coincidence backgrounds.
\item Low-energy neutrons ($\sim$\,690\,keV) were used so that low-energy nuclear recoils ($<$\,50\,\keVr) could be obtained at large neutron scattering angles and the relative angular uncertainties were reduced.
\item A small \naitl\ crystal (25\,mm cube) was used to reduce multiple scattering backgrounds.
\item A high-quantum-efficiency photomultiplier (PMT) was used to enhance light collection ($\sim$\,18\,photoelectrons/\keVee\ achieved).
\item Pulse-shape discrimination (PSD) methods were used to select neutrons events and to reject gammas and noise.
\end{enumerate}

This experiment was inspired by the success of the SCENE experiments, which measured the nuclear recoil quenching effects in liquid argon down to very low nuclear recoil energy ($\sim$10\,\keVr) using the Notre Dame facility~\cite{SCENE2013_FieldQuenching, SCENE2014_Quenching}.

\section{Experimental Setup}
\label{sec:setup}

\subsection{Overview}
The measurement of the Na quenching factor was performed at the FN Tandem accelerator at the University of Notre Dame Institute for Structure and Nuclear Astrophysics. A pulsed beam of protons from the accelerator interacted with a LiF target to produce neutrons with a nominal energy of 690\,keV at 0\textdegree\ scattering angle.
A detector consisting of an enclosed NaI(Tl) crystal and photomultiplier tube (PMT) was placed on the beam line.  The neutrons traveling in this direction could produce nuclear recoil scintillation events in the crystal and be subsequently detected by a liquid-scintillator-based neutron detector at a fixed recoil angle.  The kinematics of this interaction determined the nuclear recoil energy. 

\begin{figure}
\centering
\includegraphics[width=.5\textwidth]{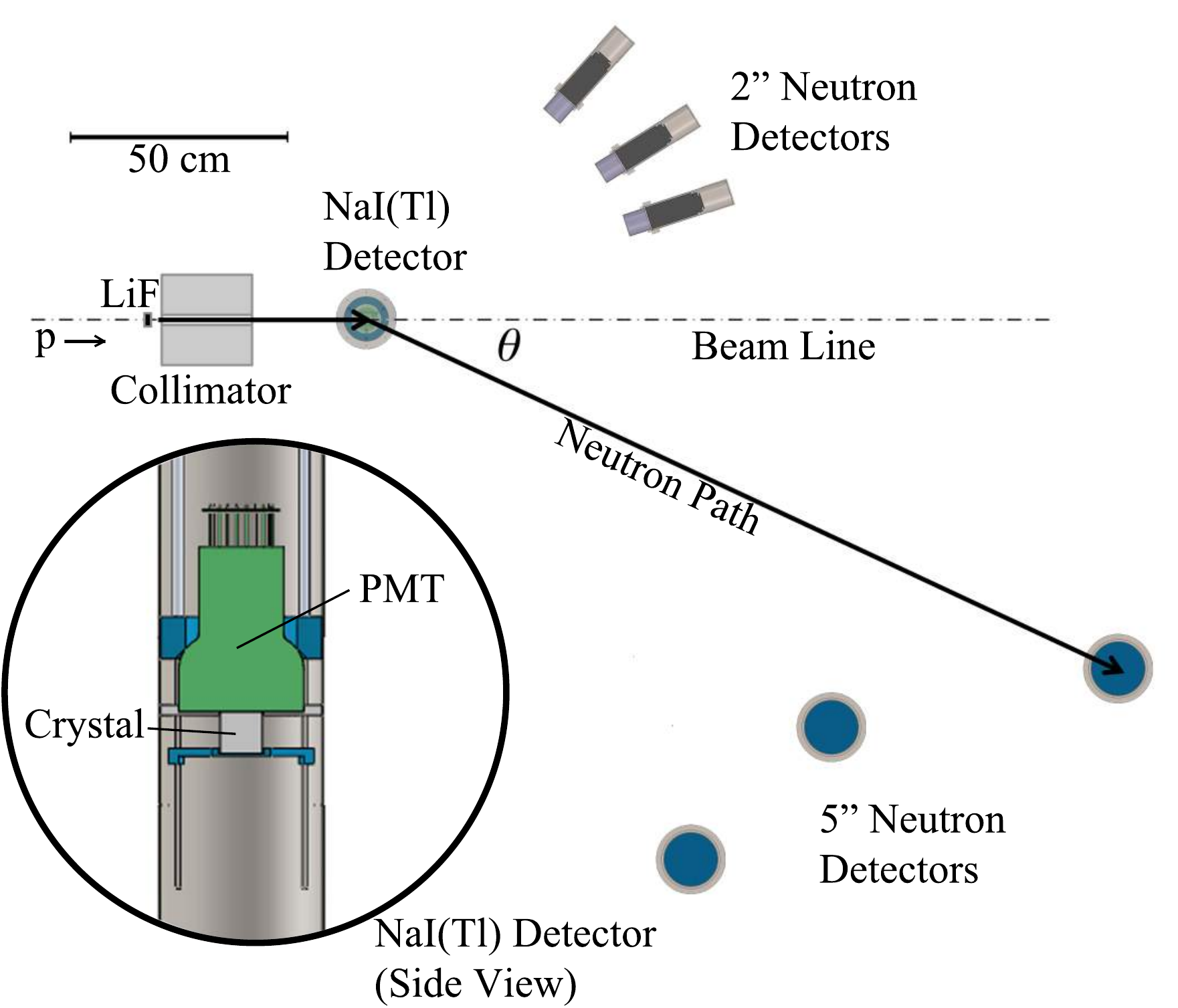}
\caption{To-scale, bird's-eye view of the experimental setup in the first position configuration with side view of the NaI(Tl) detector.  A LiF target in the beam produces neutrons when struck by the proton beam.  The neutrons travel to the NaI(Tl) detector, where they scatter off a sodium or iodine nucleus to one of the neutron detectors positioned at the relevant angle.  Two different types of neutron detectors were used, as described in the text below.  A polyethylene collimator prevents neutrons from hitting the neutron detectors directly from the LiF target.  Also shown is a side view of the NaI(Tl) detector, with the crystal shown in white and the PMT in green. }
\label{Exp:exp_setup}
\end{figure}

This information, combined with the scintillation light collected by the NaI(Tl) detector, provided a measure of the light yield of the detector for nuclear recoil events.  Calibrating this system with electron recoils of known energies allowed for a determination of the quenching factor.  We used a single NaI(Tl) detector in two positions and a stationary array of six neutron detectors, thereby measuring 12 nuclear recoil energies.  A scheme of the experimental setup in the first position configuration is shown in Figure \ref{Exp:exp_setup}.

\subsection{The Proton Beam and LiF Target}
A beam of 2.44-MeV protons was produced by an 11-MeV FN Tandem accelerator.   
These protons, incident on a LiF target, produce neutrons through the $^{7}$Li(p,n)$^{7}$Be reaction (Q-value: $-1.644$\,MeV).  Ignoring energy loss in the target, the neutrons emitted at a nominal scattering angle of 0\textdegree\ have an energy of 723\,keV.

The beam was separated into time-bunched pulses by a three-part  pulsing system with a timing resolution of 2\,ns and an intrinsic period of 101.5\,ns. 
The proton pulse selector can additionally be set to allow only one out of every $n$ pulses to pass through, effectively increasing the period.  The beam cross section is 3\,mm in diameter and is highly stable, with a variation in proton energy of around 1\,keV.  

The energy of the beam was chosen to increase the event rate while reducing the relative angular uncertainty for nuclear recoils.  Because of the finite detector size of both the NaI(Tl) detector and the neutron detectors, as well as the shape of the recoil-energy dependence on angle, the spread in the nuclear recoil energies deposited in the crystal is smaller when the neutron detectors are located at larger angles with respect to the beam. We therefore wanted the proton beam energy to produce the recoil energies of interest at relatively large scattering angles, while providing enough event rate to collect adequate statistics for the measurement.  A calculation was done to obtain the overall interaction rate in the NaI(Tl) detector at low recoil energies given the \lipn cross sections in  \cite{Burke1974_LipnBe} and the differential neutron elastic scattering cross sections in $^{23}$Na and $^{127}$I at the appropriate angles. A broad maximum in the event rate for all recoil energies was found at a proton energy of 2.44\,MeV with reasonably large scattering angles for recoil energies between 3 and 52\,keV, and therefore that energy was chosen for the measurement.


A higher pulse frequency can lead to a higher neutron event rate, but it may also cause pileups in the NaI scintillation time window.  Based on a calculation of the neutron yield, we determined that a pulse separation of 609\,ns was enough to reduce the pileup rate. As described in Section \ref{exp:sec:detectors}, the detectors were arrayed in two geometrical configurations. In the first position configuration, where the NaI(Tl) detector was 50\,cm from the LiF target, the bunching ratio for the pulser was set to 1 in 6, for an effective pulse period of 609\,ns, but was later changed to 1 in 8 after a high event rate was observed, for a pulse period of  812\,ns.  The second position configuration, with a NaI(Tl) distance of 91\,cm, had a bunching ratio of 1 in 8.  Each pulse carried $\sim10^{4}$ protons.

The LiF target was deposited on a 0.4-mm tantalum backing, which stops the proton beam and minimizes the gamma background. Incoming protons lose energy as they travel through the LiF target, leading to a broadening of the outgoing neutron energy spectrum.  The LiF target thickness was chosen to be 0.52\,mg/cm$^{2}$ in order to compromise between the event rate and the spread in the neutron energy, which both increase with thickness.  
The mean neutron energy for a target thickness of 0.52\,mg/cm$^2$ was calculated to be 690\,keV with a spread of 4\% at the full-width-half-maximum.  At that thickness, and with a bunching ratio of 1 in 8, the neutron flux was calculated to be around 300\,neutrons/s at the NaI(Tl) detector for the first position configuration and about 100\,neutrons/s in the second.  

\subsection{The Detectors}
\label{exp:sec:detectors}
The NaI(Tl) detector consisted of a 25-mm cubical NaI(Tl) crystal optically coupled to a 76-mm super-bialkali Hamamatsu R6233-100 PMT.   The crystal was grown at Radiation Monitoring Devices Inc. with high-purity Astro-grade NaI powder from Sigma Aldrich.  The small crystal size was chosen to minimize the probability of neutron multiple scattering.  The crystal and the PMT were packaged in a stainless-steel enclosure with a thin wall ($\sim$0.5\,mm) in the section  surrounding the crystal to further minimize the chance of multiple scatters.  

A high light-collection efficiency of the detector is necessary to obtain a high energy resolution and low threshold.  The PMT had a high peak quantum efficiency of $\sim$35\%.  In addition, the crystal was covered on the other five faces with highly-reflective Lumirror reflector ($>$98\% reflective above 350\,nm) additionally wrapped in several layers of PTFE tape.  No light guide was used in this experiment; the coupling  was a transparent optical gel from Cargille Labs with a refractive index of 1.52.  This arrangement allowed for a high maximum light yield of 18.2$\pm0.1$\,photoelectrons (p.e.)/\kevee, where a \kevee is a unit describing the electron-equivalent energy that would produce the same scintillation yield as a nuclear recoil with an energy of 1\,\keVr. 



The NaI(Tl) detector was placed on the beam-line axis to maximize the event rate.  The NaI(Tl) detector was placed 50\,cm from the LiF target in the first position configuration and 91\,cm in the second configuration, as described in Table \ref{tab:detectors}. The 50\,cm position was chosen in order to produce a high event rate while keeping the total angular spread, $\Delta\theta/\theta$, below 5\%.  The 91-cm position was chosen to increase the number of recoil energies explored without changing the already well-established locations of the neutron detectors.  
 
The angles chosen for the neutron detectors, between 18 and 84\,degrees, allowed for data to be collected for Na nuclear recoil energies between 3 and 52 keV.  The detector distances were chosen to maximize the event rate while maintaining a recoil energy uncertainty due to finite detector size of less than 5\%. Their positions are summarized in Table \ref{tab:detectors}. 
Two types of neutron detectors were used for the measurement: 5.1-cm x $\phi$5.1-cm Eljen \mbox{510-20x20-9/301} and 12.7-cm x $\phi$12.7-cm Eljen \mbox{510-50x50-1/301} liquid scintillator detectors.  Both types have the reflector EJ-510 and the liquid scintillator EJ-301, a xylene-based scintillator with organic fluors.  This scintillator has pulse-shape discrimination capability, which allows for the selection of events induced by desired particle types. A typical light yield response of these detectors was measured to be $\sim$1\,p.e./\kevee.

A 22-cm-diameter, 22-cm-long, cylindrical polyethylene collimator with a 2.5-cm-diameter hole was used to prevent neutrons from traveling directly from the LiF target to the neutron detectors.

\begin{table}
\begin{center}
\caption{Detector information and positions for position configurations 1 (top) and 2 (bottom). The ``Flight Distance'' for the neutron detectors (ND) is defined as the distance from the NaI(Tl) detector to the neutron detector, while for the NaI(Tl) detector it is the distance from the LiF target to the NaI(Tl) detector. The neutron detectors are cylindrical Eljen detectors with EJ-301 as the scintillator and EJ-510 as the reflector.  The detector size given is both the diameter and length of the cylinder.} 
\label{tab:detectors}
\begin{tabular}{c|c|c|c|c}
\hline
\hline
Detector	& Detector &Scattering & Recoil&  Flight  	\\
		& Size (cm) 	&Angle (deg) 	& Energy (keV)	& Distance (cm)\\
\hline
\hline
NaI(Tl) 	&2.5	& 	0	& 		&	50 	\\
		& 	& 	0	& 		&	91 	\\
\hline \hline
ND1 		&12.7&	59.1	&	29.0	&	150	\\
		&	&	74.2 	&	43.0	&	135	\\
\hline
ND2 		&12.7&	41.3 &	15.0	&	150	\\
		&	&	54.4	&	24.9	&	122	\\
\hline
ND3 		&12.7&	24.9	&	5.7	&	200	\\
		&	&	31.1	&	8.8	&	164	\\
\hline
ND4 		&5.1	&	47.9	&	19.4	&	70	\\
		&	&	84.0	&	51.8	&	52	\\
\hline
ND5 		&5.1	&	32.2	&	9.1	&	70	\\
		&	&	64.6	&	33.3	&	41	\\
\hline
ND6 		&5.1	&	18.2	&	2.9	&	70	\\
		&	&	41.1	&	14.3	&	33	\\
\hline
\hline
\end{tabular}
\end{center}

\end{table}

\subsection{Electronics and Data Acquisition}

\begin{figure}
\centering
\includegraphics[width=.4\textwidth]{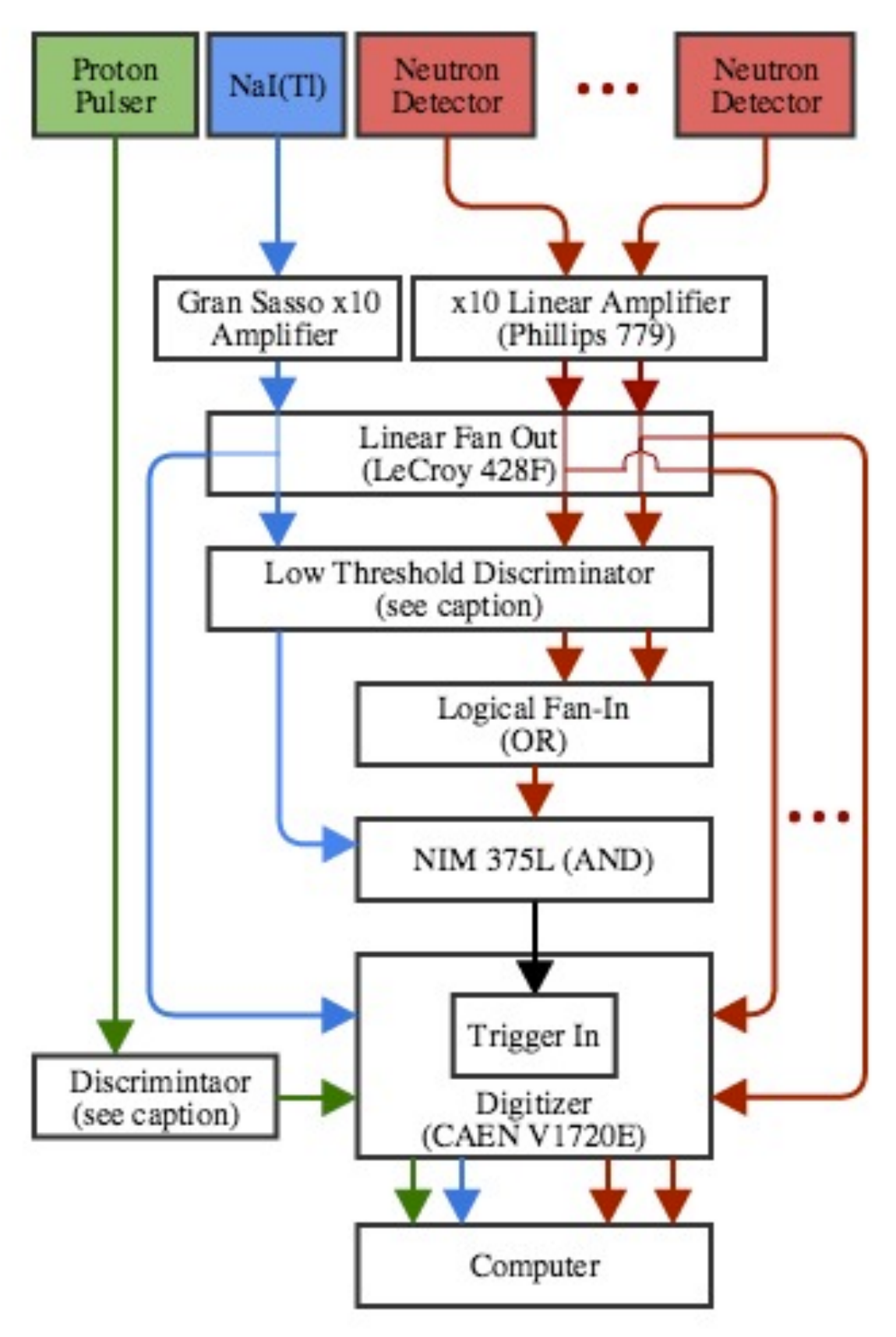}
\caption{Electronics scheme for the measurement. The NaI(Tl) signal was fed through a front-end amplifier module developed at LNGS.  The NaI(Tl) and neutron detectors with lower gain were passed through a Phillips 711 discriminator with a threshold of 10 mV, while other neutron detectors and the proton pulse selector were passed through the LeCroy 621 AL discriminator with a higher threshold equivalent to 1-2 photoelectrons in the neutron detectors. }
\label{fig:electronics}
\end{figure}

The data acquisition system needed to provide an accurate determination of the event energy and timing, as well as the particle type through pulse-shape discrimination. This could be achieved by recording the waveforms from the photomultiplier tube signals in both the NaI(Tl) and the neutron detectors, as well as the signal from the proton pulse selector, during neutron-induced scintillation events.    
The data acquisition scheme is shown in Figure \ref{fig:electronics}.  Signals from both the NaI(Tl) and the neutron detectors were amplified and fanned out.  One copy of the signals was used to produce a trigger for a CAEN V1720E digitizer module (12 bit, 250\,MS/s), while the other copy was digitized. For each trigger, a signal region of 8\,\us\  with 2\,\us\ before the trigger was digitized to ensure that the entire NaI(Tl) scintillation event was recorded (scintillation lifetime $\tau$=200-300\,ns).

The PMT signal from the NaI(Tl) detector was amplified with a x10 front-end amplifier module developed at the Laboratori Nazionali del Gran Sasso (LNGS) while the neutron detector signals were sent to a Phillips 779 x10 amplifier.  Amplified signals from both detectors were sent to a LeCroy 428F linear fan-out.  The signals to be used for the trigger first went to low-threshold discriminators (Phillips 711 and LeCroy 621 AL), whose discrimination levels were set at 1.5\,p.e. in order to reach a low energy threshold while reducing the random coincidence rate.  The discriminator outputs for the neutron detectors combined by a NIM logical fan-in whose OR output was subsequently combined with the NaI(Tl) discriminator output in a logical AND (NIM 375L).  The coincidence window for this AND logic was set to 400\,ns to conservatively include the neutron coincidence events with the longest time of flight.  Subsequent triggers within the acquisition window were discarded.  The signal from the pulsed proton beam was not used in the trigger, but was recorded for off-line analysis.  Due to the degradation of the proton pulse selector signal in long transmission lines, the signal was fed through a discriminator before being digitized. 

 A basic online analysis was performed to show the scintillation waveforms, the coincidence event rate, and the time-of-flight spectra for all neutron detectors. The waveform data from all channels were saved to disk in a binary format in real time, to be used for offline analysis, as discussed in Sec. \ref{sec:analysis}.


\subsection{Measurement Summary}
Data were collected in the first position configuration for 26 hours and in the second position configuration for 20 hours, giving approximately 1,000--4,000 coincidence events per energy. 

Calibrations of the light yield of the NaI(Tl) detector were taken in-run by observing the 57.6\,keV gamma ray that comes from the first excited state of $^{127}$I, which can be induced through inelastic scattering of the neutrons. The light yield of this detector was initially measured to be 18.2\,p.e./\kevee, but decreased over the course of the measurement  to 13.7\,p.e./\kevee\ due to some degradation of the crystal from moisture exposure, and possibly also a degradation of the optical coupling.  The in-run calibration compensated for the loss of light yield in the calculation of an energy spectrum for the nuclear recoil events. 

Separate calibration runs with $^{133}$Ba and $^{241}$Am sources also observed this light-yield degradation.  However, a 
few-percent systematic difference in the light yield between the two measurements was observed; the source calibrations  showed a lower light yield than the real-time calibration with $^{127}$I.  One potential effect is the skewing of the peak due to the energy of the iodine recoil itself, but this effect was estimated to be at or below 1\%.  Another potential reason for this difference stems from the position distribution of the scintillation events; the gammas from the first excited state of $^{127}$I are evenly distributed throughout the crystal, whereas the gammas from the external sources will interact within a few mm of the crystal edge.  This effect can cause a systematic decrease in the light yield, as the light yield may be position dependent. 
The in-beam 
measurement of the 57.6\,keV peak was used to calibrate our detector performance in our analysis.

\begin{figure}
\centering
\includegraphics[width=.48\textwidth]{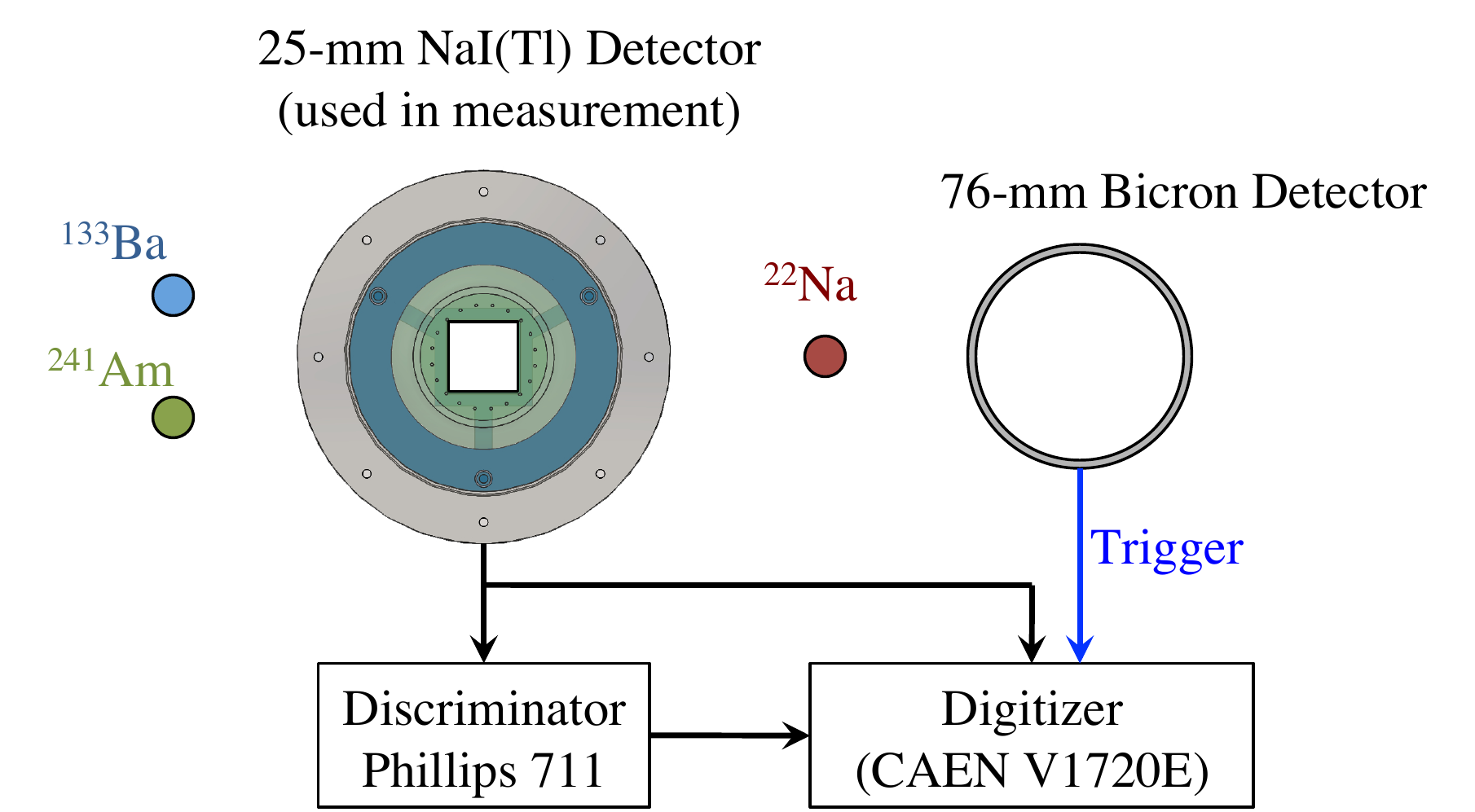}
\caption{Setup for the trigger efficiency measurement of the 25-mm NaI(Tl) detector.  A Na-22 source was placed between the NaI(Tl) detector and a separate 76-mm Bicron NaI(Tl) detector so that the back-to-back 511\,keV gamma rays could be detected by both detectors.  The signal was fed both directly into the digitizer and also through the discriminator, while the Bicron detector was used as a trigger to reduce the acquisition of data with no pulses.  Additional sources were placed near the 25-mm detector to increase the event rate.}
\label{exp:trig_eff}
\end{figure}

After the data were collected for the measurement of the quenching factor, a separate run was conducted to measure the trigger efficiency of the NaI(Tl) detector at low energies.  The setup is shown in Figure \ref{exp:trig_eff}. A Bicron 76-mm NaI(Tl) detector was set up at a reasonable distance away from the 25-mm NaI(Tl) detector.  A $^{22}$Na source was placed directly between the two detectors, which produced back-to-back 511-keV gamma rays.  $^{133}$Ba and $^{241}$Am sources were also put near the 25-mm detector opposite the 76-mm detector to increase the event rate.  The 76-mm detector was used as a trigger for the measurement with a high threshold.  The 25-mm detector signal from the discriminator as well as directly from the amplifier were fed into the digitizer to measure the efficiency of the trigger for low energy recoils. 

\section{Data Analysis}
\label{sec:analysis}

As discussed in Section~\ref{sec:setup}, data from 12 neutron-scattering angles, corresponding to 12 different nuclear recoil energies, were collected in this measurement. The nuclear recoil signals in the \naitl\ crystal were selected using time-of-flight (TOF) and pulse-shape-discrimination (PSD) cuts. Their energy spectra were then compared to the predicted recoil-energy spectra, produced by Monte Carlo simulations, to evaluate the energy-dependent nuclear recoil quenching factors.  We report the analysis of sodium recoil quenching effects in an energy window of $\sim$\,3\,\keVr\ to $\sim$\,52\,\keVr, which covers the \dl\ region of interest. Iodine recoils were not observed in the measurements due to their low recoil energies and larger quenching effects; limits were set on the iodine quenching factors.

\subsection{Data Processing}
\label{sec:dataprocess}

The raw data acquired in the measurements contain the waveforms of the \naitl\ detector, the waveforms of 6 liquid scintillator neutron detectors, and a periodic pulse-selector signal from the proton accelerator, which relates to the proton-on-target (POT) time with a constant offset. The waveform baselines were first subtracted using a drifting-baseline-finding algorithm, which was tuned to suppress low-frequency electronic noise while preserving high-frequency scintillation pulse signals. Individual pulses with an amplitude $\gtrsim$\,0.2 photoelectrons  were tagged and further processed for the analysis. For each pulse that  contributed to a coincidence trigger, all following pulses within 4\,\us\ were clustered together as one scintillation event for energy evaluation. The 4\,\us\ time window was chosen to contain the full scintillation signals of the highest-energy events considered in this analysis.

\begin{figure*}[t]
\centering
\begin{minipage}[c]{0.45\textwidth}
\centering
\includegraphics[angle=0,height=0.7\textwidth]{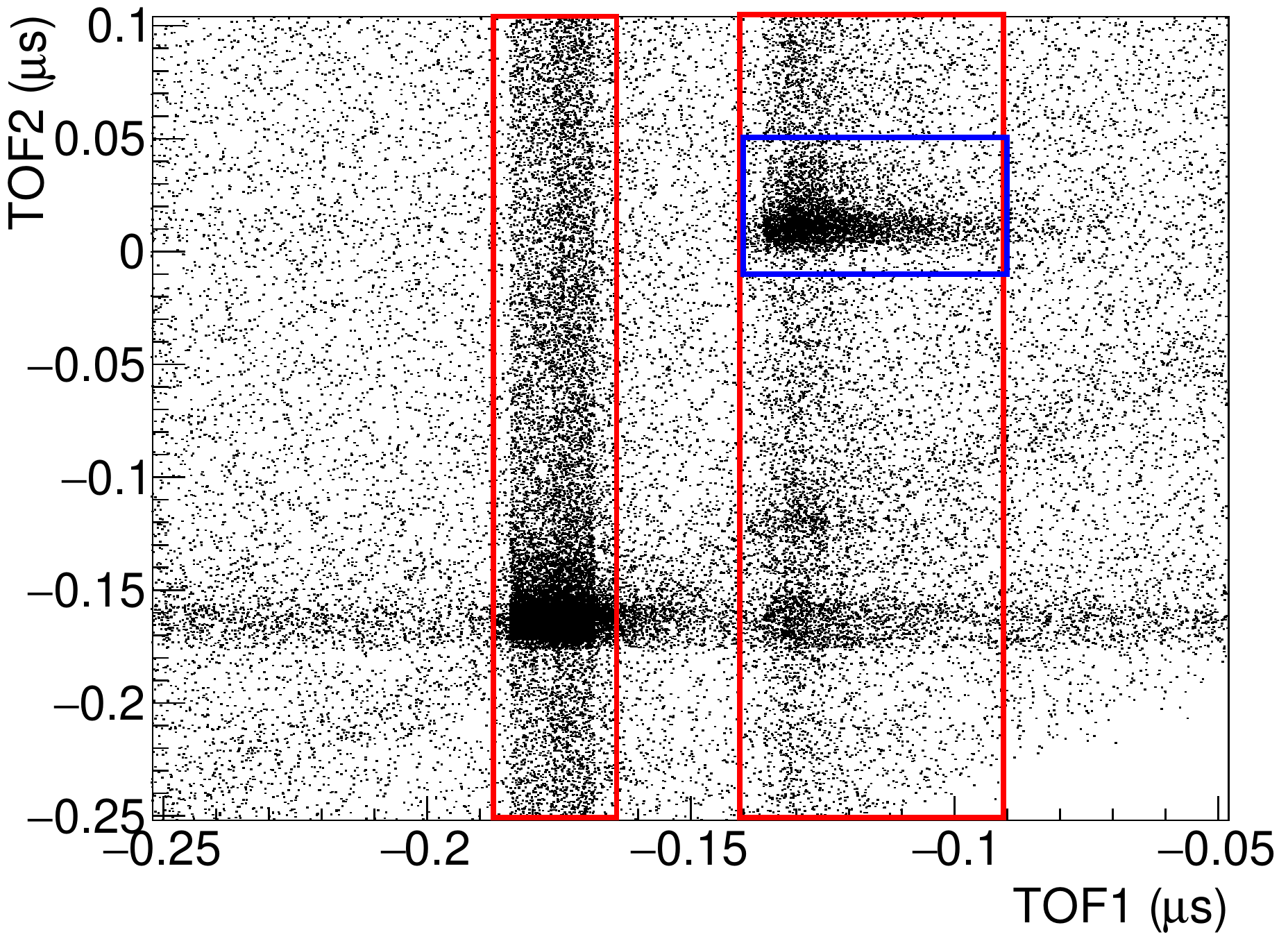}
\end{minipage}
\begin{minipage}[c]{0.45\textwidth}
\centering
\includegraphics[angle=0,height=0.69\textwidth]{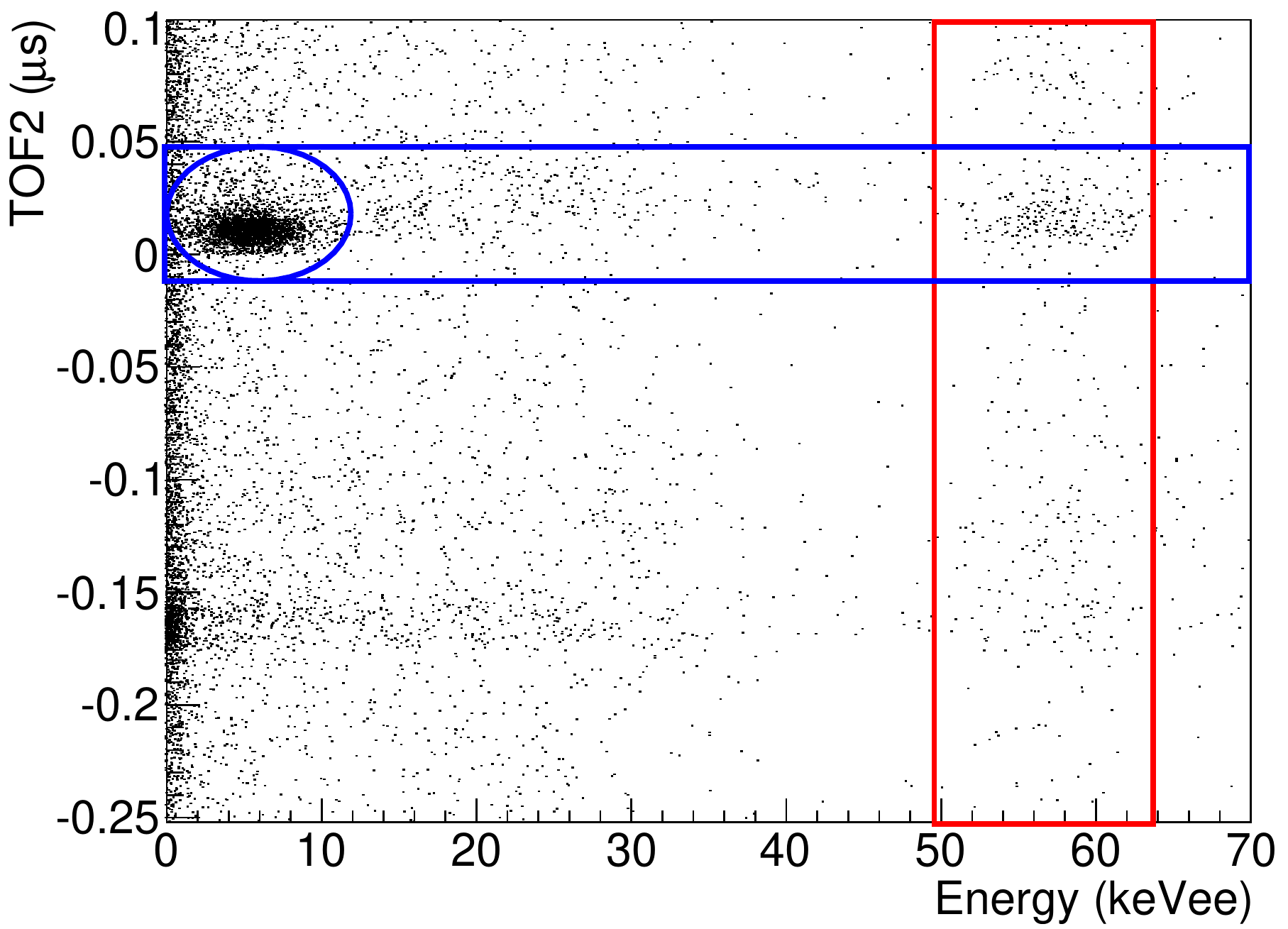}
\end{minipage}
\caption{\textbf{Left:} The double TOF spectrum for coincidence events containing $\sim$\,29\,\keVr\ Na recoils. Gammas (first vertical band) and neutrons (second vertical band) from the LiF can be separated by TOF1, defined as the TOF from LiF to \naitl; neutron scattering off \naitl\ can be further selected (blue box) using TOF2, defined as the total TOF from the LiF to the liquid scintillator neutron detectors. \textbf{Right:} The energy distribution of the neutron induced events from the second band in the figure on the left. The blue box contain the neutron scattering events. Neutron-induced nuclear recoils are marked in the oval and the \iseven\ excitation events are shown in the vertical band.}
\label{fig:tof}
\end{figure*}

For every coincidence event, we first tried to identify which neutron detector contributed to the trigger by comparing the pulse arrival times with the coincidence trigger time. If the signals from the responsible neutron detector and the \naitl\ detector satisfied certain event selection criteria, the \naitl\ signal is kept for the quenching factor analysis and the neutron detector position is used to determine the neutron scattering angle. The most important event selection criteria in the analysis is the cut on the time of flight (TOF), or the time difference between the pulse arrival times in different detector channels. For the neutron energy used in this measurement ($\sim$\,690\,keV, corresponding to a speed of $\sim$\,1\,cm/ns), the time required for the neutrons to travel from the \naitl\ detector to the neutron detectors was at the level of $\sim$\,50\,-\,200\,ns, depending on the detector positions. This well-defined time correlation made the neutron events distinct from the nearly instantaneous gamma coincidence background and random coincidence backgrounds. Furthermore, the pulsed neutron beam also made it possible to calculate and cut on the TOF between the LiF target and the \naitl\ detector, which further suppressed the gamma ray background generated by the proton beam.

In this analysis, TOF1 was defined as the difference between the arrival times of the \naitl\ signal and the proton pulse signal, and TOF2, or the total TOF, was defined as the difference between the arrival times of the neutron detector signal and the proton pulse signal. In a real neutron-induced coincidence event, TOF1 represents the time (with a constant offset) required for the neutron to travel from the LiF target to the \naitl\ detector, and TOF2 corresponds to the time (with a similar constant offset) required for the neutron to travel from the LiF target, to scatter off the \naitl\ detector, and to be recorded by a liquid scintillator neutron detector.

Figure~\ref{fig:tof} (left) shows the double TOF spectrum (TOF2 vs. TOF1)  for the coincidence events between the \naitl\ detector and a neutron detector containing $\sim$\,29\,\keVr\ Na recoils. The first and second vertical bands, respectively, correspond to the gammas and neutrons that were produced by the proton beam and recorded by the \naitl\ detector. Because the gamma-ray flight time from the target to the \naitl\ detector is only a few nanoseconds, the time separation between the two vertical bands provided an estimate of the neutron TOF, which was confirmed by direct calculations. The experiment was designed in a way that the gamma and neutron TOF bands were sufficiently separated, so conservative TOF cuts could be used to efficiently reject the gamma background with little impact on the neutron events. Events in the horizontal band with TOF2 slightly below -0.15\,$\mu$s were identified to be the beam-induced gamma rays that directly hit the neutron detector in coincidence with a random \naitl\ scintillation event. The diagonal band with TOF2\,$\approx$\,TOF1 was attributed to simultaneous scintillations in the \naitl\ detector and in the neutron detector from environmental radioactivity such as high-energy gammas or comic-ray showers. Combining the analysis using both TOF1 and TOF2, the blue box in Figure~\ref{fig:tof} (left) was identified to contain the desired neutron coincidence events where a neutron scattered off the \naitl\ detector and then got recorded in the neutron detector. 

In addition to nuclear recoils, neutron interactions with \naitl\ also produced nuclear excitations via inelastic scattering. As shown in Figure~\ref{fig:tof} (right), the 57.6\,keV \iseven\ excitation gamma rays were observed (vertical band in the plot) in all neutron induced \naitl\ scintillation energy spectra. As discussed in Section~\ref{sec:setup}, these gammas were used to provide an in-run energy calibration for this measurement; they also provided a way to monitor and correct the degradation of the \naitl\ light yield observed between the runs. It was estimated that this calibration introduced $\sim$\,1\,\% uncertainty in the energy scale due to the iodine recoils accompanying the 57.6\,keV gamma rays, and the time-dependent light yield correction further introduced a 1.5\% uncertainty. The quenching factors to be reported in this paper are all normalized to the scintillation efficiency of \naitl\ under 57.6\,keV gamma excitations. Due to a possible non-linearity of the \naitl\ scintillation output~\cite{knoll_raddetection, DAMA2008_apparatus}, the evaluated nuclear-recoil quenching-factor values may depend on the gamma calibration point, and the results from different measurements need to be appropriately scaled for direct comparison.

\begin{figure*}[t]
\centering
\begin{minipage}[c]{0.45\textwidth}
\centering
\includegraphics[angle=0,height=0.68\textwidth]{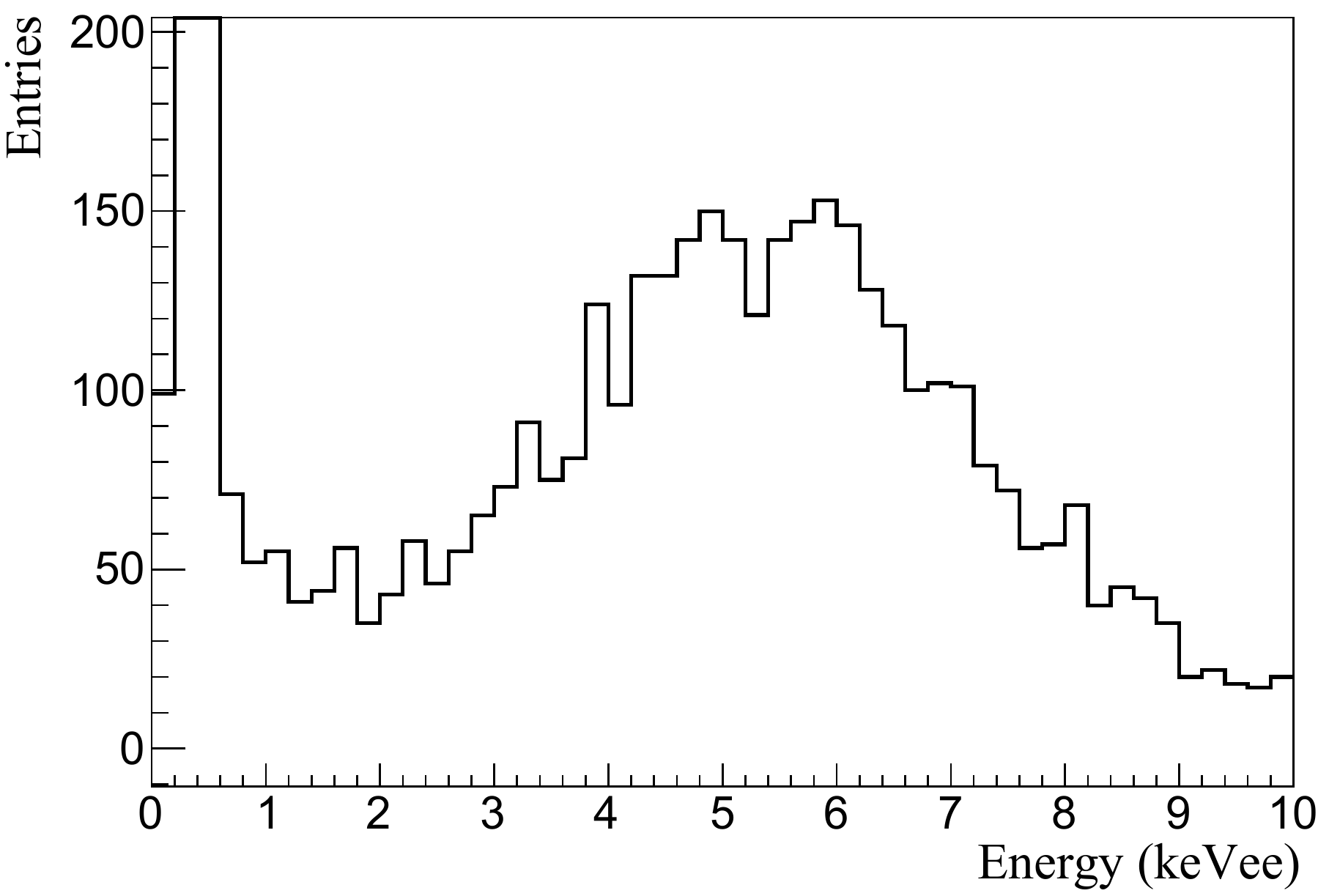}
\end{minipage}
\begin{minipage}[c]{0.45\textwidth}
\centering
\includegraphics[angle=0,height=0.68\textwidth]{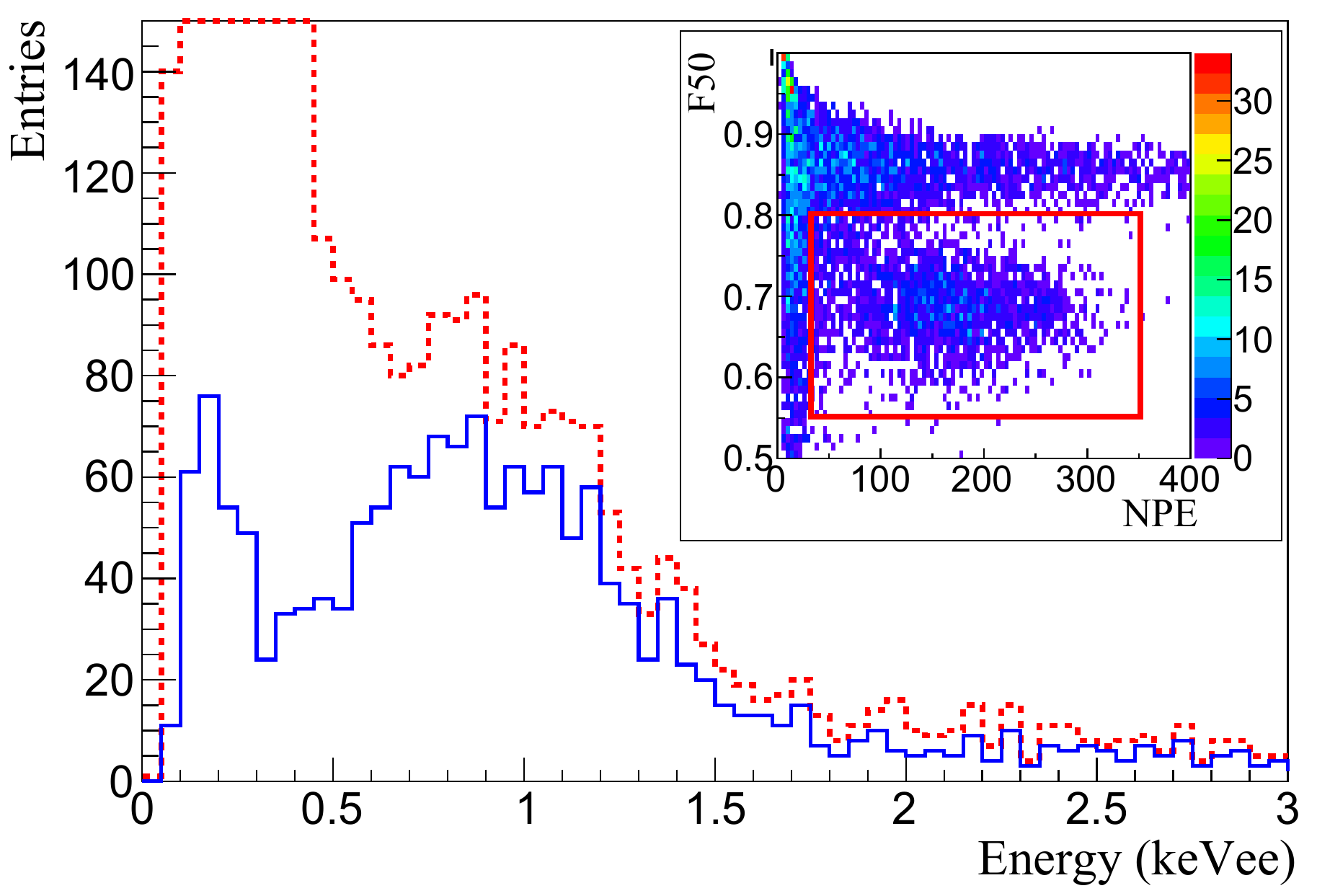}
\end{minipage}
\caption{Examples of observed Na recoil energy spectra after the conservative TOF cuts. \textbf{Left:} Energy spectrum of $\sim$\,29\,\keVr\ Na recoils (selected from Figure~\ref{fig:tof}). Only cuts on TOF1 and TOF2 were applied. \textbf{Right:} The energy spectrum of $\sim$\,5.7\,\keVr\ Na recoils with (solid blue) and without (dotted red) PSD cuts on the liquid scintillator detector signals. The insert figure shows the PSD parameter F50 vs. the pulse integral (in the unit of number of photoelectrons, or NPE) in the neutron detector. Note that PSD cuts were only necessary for data below the \dl\ energy threshold of 2\,\keVee.}
\label{fig:naspect}
\end{figure*}

With the neutron events selected using TOF cuts and the energy scale calibrated with \iseven\ excitation gammas, the nuclear recoil energy spectra were extracted for all 12 neutron scattering angles. The single-scattering Na recoil peaks could be resolved clearly in the relatively high-energy regions ($>$10\,\keVr), as illustrated in Figure~\ref{fig:naspect} (left), which shows the energy spectrum of $\sim$\,29\,\keVr\ Na recoil events selected from Figure~\ref{fig:tof}. However, the energy spectra of relatively low-energy Na recoils ($<$10\,\keVr) were contaminated by noise from random coincidence backgrounds, as shown in Figure~\ref{fig:naspect} (right). Fortunately, liquid scintillators have a scintillation pulse shape that depends on the particle type~\cite{knoll_raddetection}, a fact that we exploited to reject gamma backgrounds. As shown in the insert figure of Figure~\ref{fig:naspect} (right), a conservative cut on F50, defined as the fraction of the pulse integral in the first 50\,ns, was found to be sufficient in rejecting the low energy background events. A similar F50 cut was applied to the \naitl\ detector signals to reject fast Cerenkov light in the \naitl\ crystal and fast noise signals in the electronics; typical \naitl\ scintillations have a lifetime of $\sim$\,200\,-\,250\,ns and the bias from this conservative F50 cut was negligible. We comment that the PSD cuts were only necessary for the analysis at the lowest energies ($<$\,1.5\,\keVee), below the \dl\ energy threshold of 2\,\keVee.

With the double-TOF cuts and the PSD cuts at low energies, the spectral peaks from single neutron-Na scattering were identified at all neutron scattering angles except for the 2.9\,\keVr\ recoils. We comment that all cuts used in this analysis were carefully chosen to be overly conservative so the analysis of neutron-induced events should not be biased. Thanks to the application of multiple event-selection criteria including double-TOF cuts and double-PSD cuts, the conservative cuts, when combined, were demonstrated to be sufficiently efficient in suppressing background events that were not due to neutron scattering coincidences.

\subsection{Quenching Factor Evaluation}

To evaluate the Na recoil quenching factors, the recorded nuclear recoil spectra need to be compared to the expected recoil spectra. In principal this is straightforward to calculate with the given kinematics. However, a number of effects could introduce uncertainties, including the proton energy loss in the LiF target, the energy dependence of the \lipn\ reaction, the finite size of the \naitl\ detector, and the finite size and angular positions of the liquid scintillator neutron detectors. In this analysis, we employed  Monte Carlo techniques to calculate the energy spectra of sodium and iodine nuclear recoils, taking into account these effects.

\begin{figure}[h!]
\centering
\includegraphics[width=0.4\textwidth]{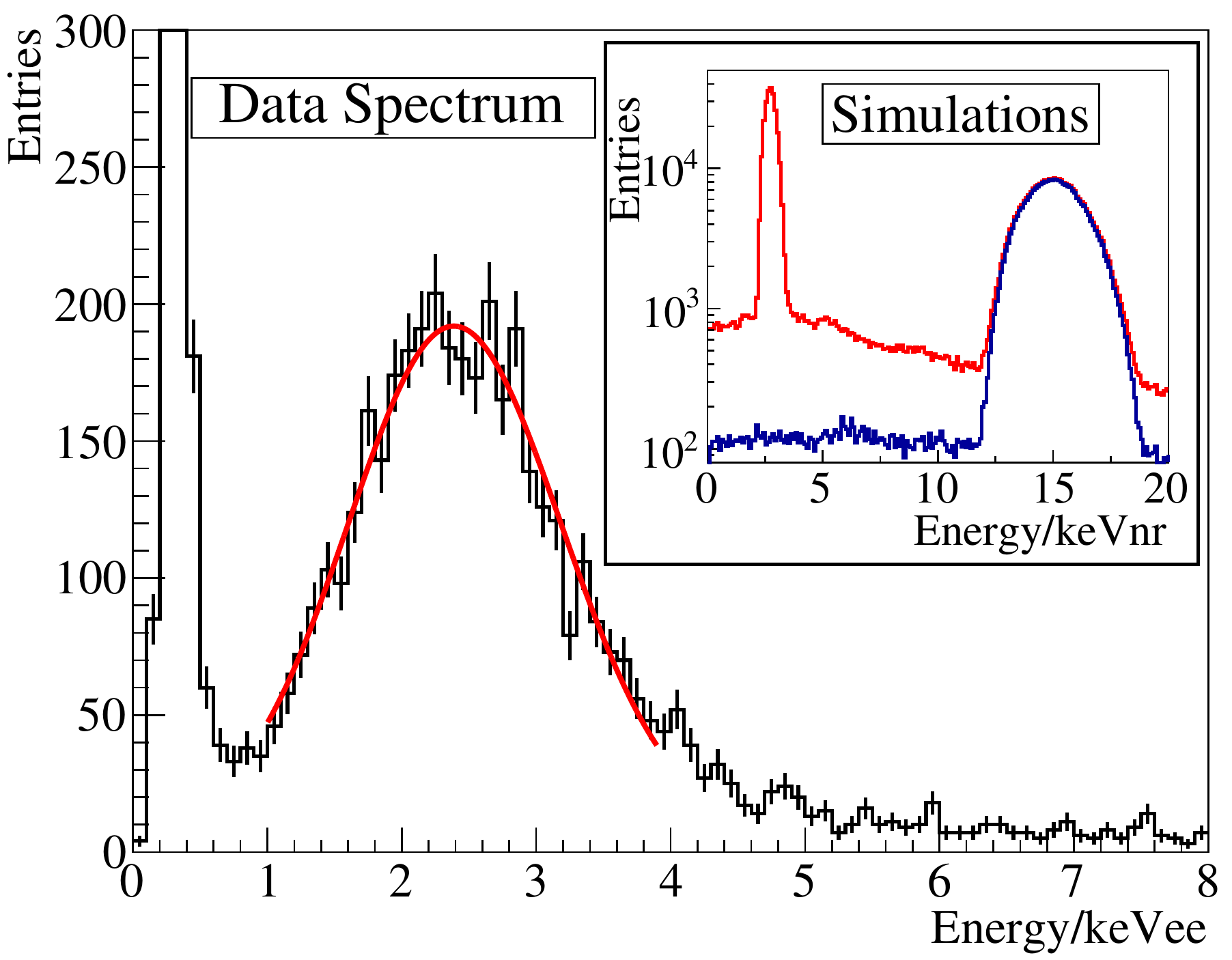}
\caption{Spectral fit of $\sim$\,15.0\,\keVr\ Na recoil events with the Monte Carlo-simulated spectrum with a Gaussian spread. The measured recoil spectrum is shown in black and the fit function is shown in red. \textbf{Insert:} The Monte Carlo-simulated nuclear recoil energy spectra for the corresponding neutron scattering angle. The blue spectrum shows the energy of single-scattering sodium recoil events; the red also includes iodine recoils and multiple scattering events in \naitl. The background rate of iodine recoils and multiple scatterings is $\sim$\,20 times lower than that of single-scattering sodium recoils in the relevant energy region. The peak around 3\,\keVr\ in the insert figure corresponds to single-scattering iodine recoils, which fell below the trigger threshold in the measurement.}
\label{fig:simufit}
\end{figure} 

The Geant4 package~\cite{geant4} (version 4.9.6.p3) was used to simulate the neutron interactions in the experimental setup. The beam-produced neutrons were generated using the  \lipn\ cross-section data measured by Burke et al.~\cite{Burke1974_LipnBe}, with the proton energy loss in the LiF target appropriately simulated. Thanks to the small size of the \naitl\ crystal used in this measurement (a cube with 25-mm sides), single-scattering nuclear recoils dominated the simulated energy spectrum, as illustrated in the insert of Figure~\ref{fig:simufit}, while multiple-scattering events contributed a small, featureless, near-flat background. 

The Na-recoil quenching factors were extracted by fitting the observed nuclear recoil spectra to the Monte Carlo-simulated energy spectra convolved with a Gaussian spread. The maximum-likelihood algorithm was used to appropriately handle the fit in the low statistic data bins, and all fits returned reasonable $\chi^2$ values. An example of the spectral fit is shown in Figure~\ref{fig:simufit}, along with the simulated energy spectrum at the corresponding neutron detector configuration (the insert of Figure~\ref{fig:simufit}). The systematic uncertainty in the fit process was evaluated by varying the fit range and comparing the fit results.

\begin{figure}[h!]
\centering
\includegraphics[width=0.45\textwidth]{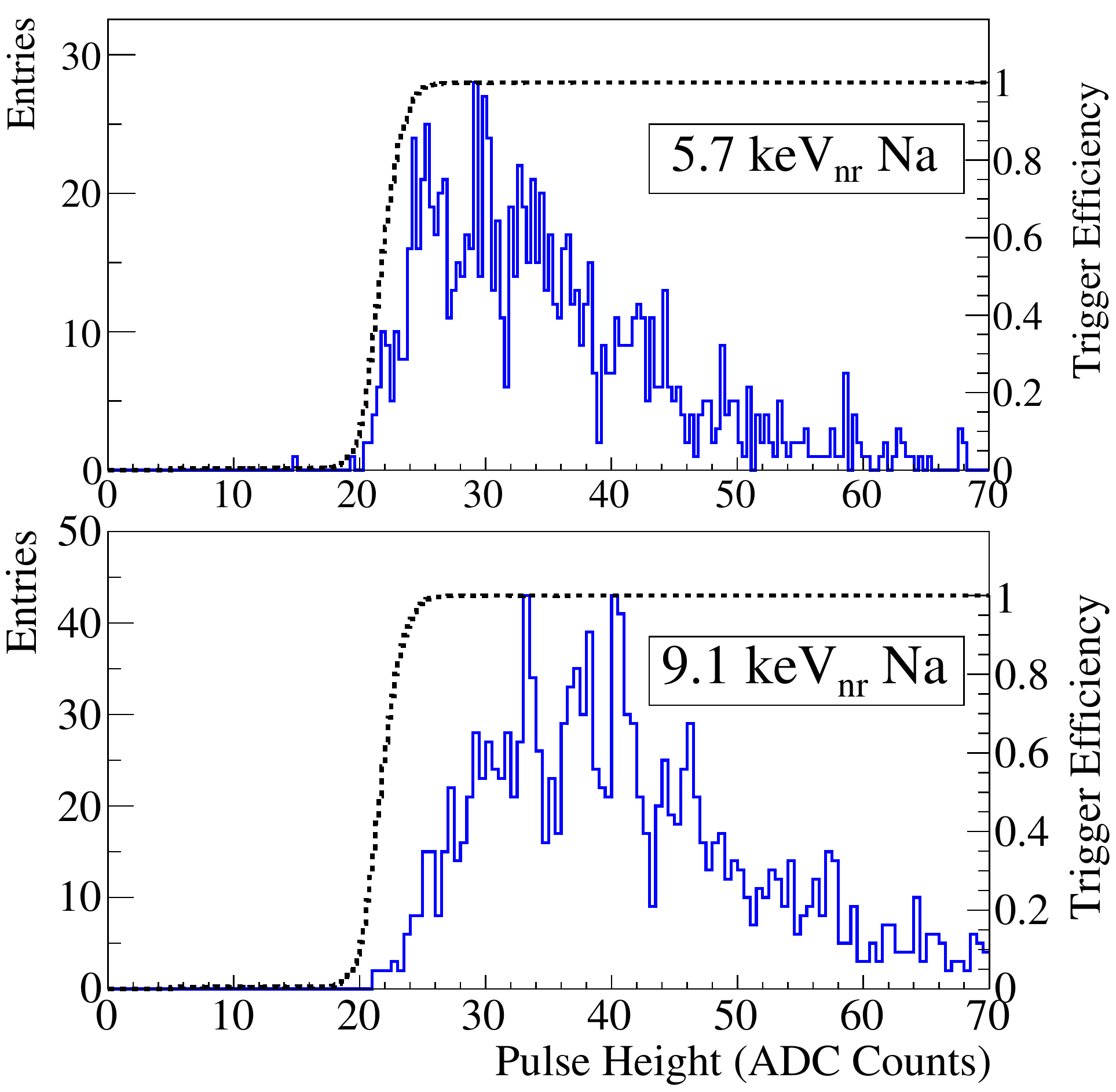}
\caption{The pulse-height spectra of 5.7\,\keVr\ Na recoil events (top) and 9.1\,\keVr\ Na recoil events (bottom). The measured pulse-height spectra are shown in blue, and the independently measured trigger efficiency is shown in black dotted line (efficiency scale is shown with the axis on the right).}
\label{fig:trigeff}
\end{figure}

Based on the trigger-efficiency measurement described in Section~\ref{sec:setup}, we found that the Na-recoil events at low energies ($<$\,10\,\keVr) were subject to loss of triggers due to the small pulse heights falling below the discriminator threshold.  As shown in Figure~\ref{fig:trigeff}, the trigger efficiency drops dramatically around 20 - 25 digitizer counts, corresponding to $\sim$\,1.5 photoelectron amplitude. It was estimated that approximately 50\% of the \naitl\ scintillation events at 0.65\,\keVee\ were lost due to the threshold effect and the loss was more significant at lower energies. This threshold effect slightly biased the low-energy tail of the 9.1\,\keVr\ Na recoil data, as illustrated in Figure~\ref{fig:trigeff}, but impacted the 8.8\,\keVr, 5.7\,\keVr\ and 2.9\,\keVr\ Na recoil data to a larger extent. 

Therefore, we adopted a different analysis strategy at these energies. First, the measured trigger-efficiency curve was used to restore the loss of triggers for events with $>$\,10\% trigger efficiency. If the peak of the pulse-height spectrum could be resolved after this correction (5.7\,\keVr, 8.8\,\keVr\ and 9.1\,\keVr), the peak position of the pulse-height spectrum was used to estimate the peak position of the pulse-integral (energy) spectrum. The pulse height/integral correlation was obtained by investigating the pulse-energy distribution at different pulse-height values. This method was confirmed to yield the correct peak energy in tests with higher-energy recoils that did not suffer trigger loss. The quenching factors were then calculated by comparing the peak positions of the observed energy with that of the predicted energy. For the Na recoils of the lowest energy (2.9\,\keVr), the pulse-height spectrum could not be effectively restored, so an upper limit of the quenched energy (and the corresponding quenching factor) was extracted.
\newline
\begin{table}[h!]
\caption{Summary of the Na quenching factors measured in this work. The angles are calculated using the central positions of the detectors; the energies reported are the peak values and widths. Quenching factors were evaluated by spectral fits between observation and simulation above 10\,\keVr\ and by comparing the peak energy positions at lower energies after correcting for the trigger-efficiency loss, as described in the text.}
\label{tab:quenchtable}
\begin{center}
\begin{tabular}{ c | c | c | c }
\hline
\hline
Scattering             & Sim. Na recoil  & Observed recoil & Quenching\\
angle ($^{\circ}$) & energy (\keVr) & energy (\keVee) & factor\\
\hline
\hline
18.2 & 2.9 $\pm$ 0.7 &  $<$0.65  & $<$ 0.22 \\
\hline
24.9 & 5.7 $\pm$ 0.7 &  0.76 $\pm$ 0.4 & 0.133 $\pm$ 0.018 \\
\hline
31.1 & 8.8 $\pm$ 1.2 &  1.13 $\pm$ 0.5 & 0.129 $\pm$ 0.014 \\
\hline
32.2 & 9.1 $\pm$ 1.2 &  1.46 $\pm$ 0.5 & 0.162 $\pm$ 0.012 \\
\hline
41.1 & 14.3 $\pm$ 2.4 &  2.21 $\pm$ 0.9 & 0.159 $\pm$ 0.019 \\
\hline
41.3 & 15.0 $\pm$ 1.4 &  2.36 $\pm$ 0.8 & 0.160 $\pm$ 0.010 \\
\hline
47.9 & 19.4 $\pm$ 1.6 &  3.21 $\pm$ 1.0 & 0.168 $\pm$ 0.009 \\
\hline
54.4 & 24.9 $\pm$ 2.4 &  4.10 $\pm$ 1.5 & 0.171 $\pm$ 0.010 \\
\hline
59.1 & 29.0 $\pm$ 1.9 &  5.36 $\pm$ 1.9 & 0.188 $\pm$ 0.008 \\
\hline
64.6 & 33.3 $\pm$ 2.8 &  6.19 $\pm$ 2.1 & 0.191 $\pm$ 0.011 \\
\hline
74.2 & 43.0 $\pm$ 2.2 &  8.53 $\pm$ 2.7 & 0.204 $\pm$ 0.008 \\
\hline
84.0 & 51.8 $\pm$ 2.6 &  10.59 $\pm$ 4.5 & 0.207 $\pm$ 0.010 \\
\hline
\hline
\end{tabular}
\end{center}
\end{table}

The Na-recoil quenching results are summarized in Table~\ref{tab:quenchtable}. Due to an asymmetry in the energy spectra, the fitted quenching factor values differ slightly from the estimates based on the peak positions of the spectra, and this uncertainty was included in the peak-comparison analysis at low energies. In addition to the uncertainties from the spectral fits and peak-comparison analysis, the results also include the 1.5\% uncertainty from the gamma-ray calibration, and 3-12\% uncertainty from the detector position measurements, which varied with neutron scattering angles and distances between detectors.

Although this measurement was designed to study the quenching effect of low-energy sodium recoils, data acquired with large scattering angles could also contain iodine recoils of up to 10\,\keVr\ based on kinematic calculations and simulations. However, we did not observe significant evidence for iodine recoils with the expected rate above 0.65\,\keVee, in which region we have over 50\% trigger efficiencies. Therefore, we have set an upper limit of 0.065 ($>$\,3\,$\sigma$) for the iodine quenching factor at 10\,\keVr. We note that DAMA uses iodine quenching value of 0.09~\cite{DamaQuench1996}, and Collar measured a much lower value of $\sim$0.05~\cite{Collar2013}.
   
\section{Results and Discussions}
\label{sec:discussions}

\begin{figure}[h!]
\centering
\includegraphics[width=0.45\textwidth]{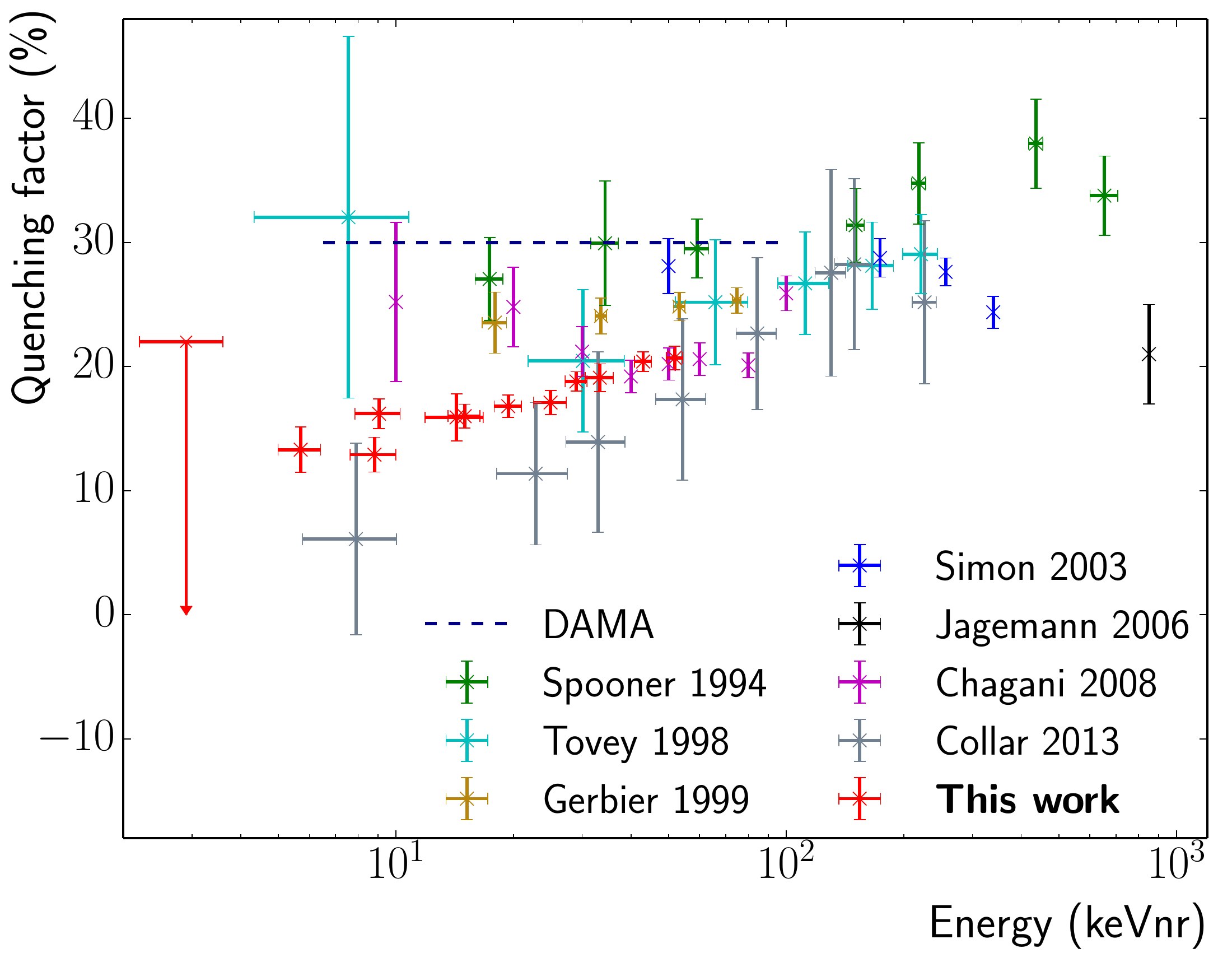}
\caption{The Na-recoil quenching factors measured in this work in comparison with previous results~\cite{Spooner1994, Tovey1998, Gerbier1999, Jagemann2006, Simon2003, Chagani2008, Collar2013}.  With much improved accuracy, this measurement approximately agrees with previous measurements at high energies and may reconcile the low-energy discrepancy. We comment that these measurements were calibrated to different gamma ray energies, and this direct comparison of quenching factors may include an uncertainty from the possible non-linearity of \naitl\ scintillation~\cite{knoll_raddetection, DAMA2008_apparatus}.}
\label{fig:qfplot}
\end{figure}

By using low-energy, pulsed neutrons and combining TOF and PSD methods, we obtained the most accurate measurement of Na-recoil quenching factors in \naitl\ to date. This result differs significantly from the measurements of DAMA, as illustrated in Figure~\ref{fig:qfplot}. At high energies, our measurement approximately agrees with previous reports, but the quenching factor values are found to drop significantly as the recoil energy decreases, similar to what was observed in the Collar measurement~\cite{Collar2013}. At low energies, our measurement falls between, and is statistically consistent with, the recent measurements by Collar and Chagani~\cite{Chagani2008}. This energy-dependent quenching effect may be explained by the increasing ionization density along the Na recoil tracks at low energies, which has a large nuclear component but small electronic component; it is usually believed that the former leads to heat generation and the latter causes scintillation~\cite{Chagani2008}.

The Na-recoil quenching factor decreases from $\sim$\,0.19 at 6\,\keVee\ to $\sim$\,0.15 at 2\,\keVee, which means that the \dl\ modulation signal (most notable at 2 - 6\,\keVee), if attributed to Na recoils, occurs in the energy window of 13\,\keVr\,-\,32\,\keVr\footnote{Here we have assumed that our quenching factors calibrated to the 57.6\,keV gammas can be directly applied to the \dl\ data because \dl\ uses gamma rays of a similar energy (59.5\,keV) for regular calibrations.} instead of the 7\,\keVr\,-\,20\,\keVr\ window, as was previously thought. Due to the exponentially decreasing characteristic of a WIMP scattering spectrum, the WIMP explanation of the \dl\ signal would require the WIMPs to have larger masses and/or larger interaction cross sections with nucleons than those assumed in previous \dl\ interpretations. 

\begin{figure}[!htp]
\centering
\begin{minipage}[c]{0.45\textwidth}
\centering
\includegraphics[angle=0,width=0.95\textwidth]{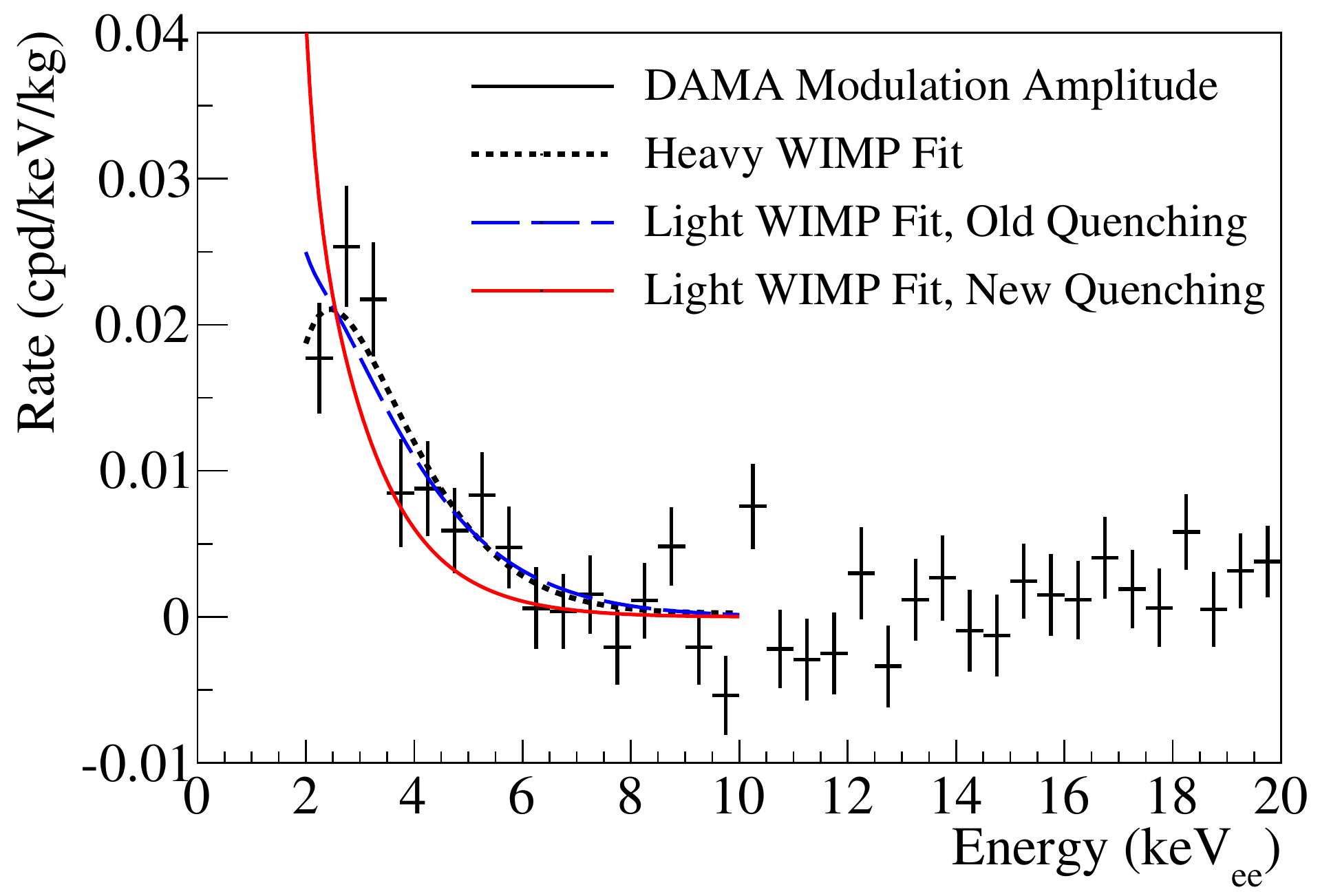}
\caption{Spectral fits of the \dl\ modulation amplitudes to the standard WIMP model (spin-independent only) with the new Na quenching factors in comparison with the fits with the old values. The heavy-WIMP fit does not change, but the low-mass-WIMP picture no longer fits the data well.}
\label{fig:fit_spect}
\end{minipage}
\begin{minipage}[c]{0.45\textwidth}
\centering
\includegraphics[angle=0,width=0.95\textwidth]{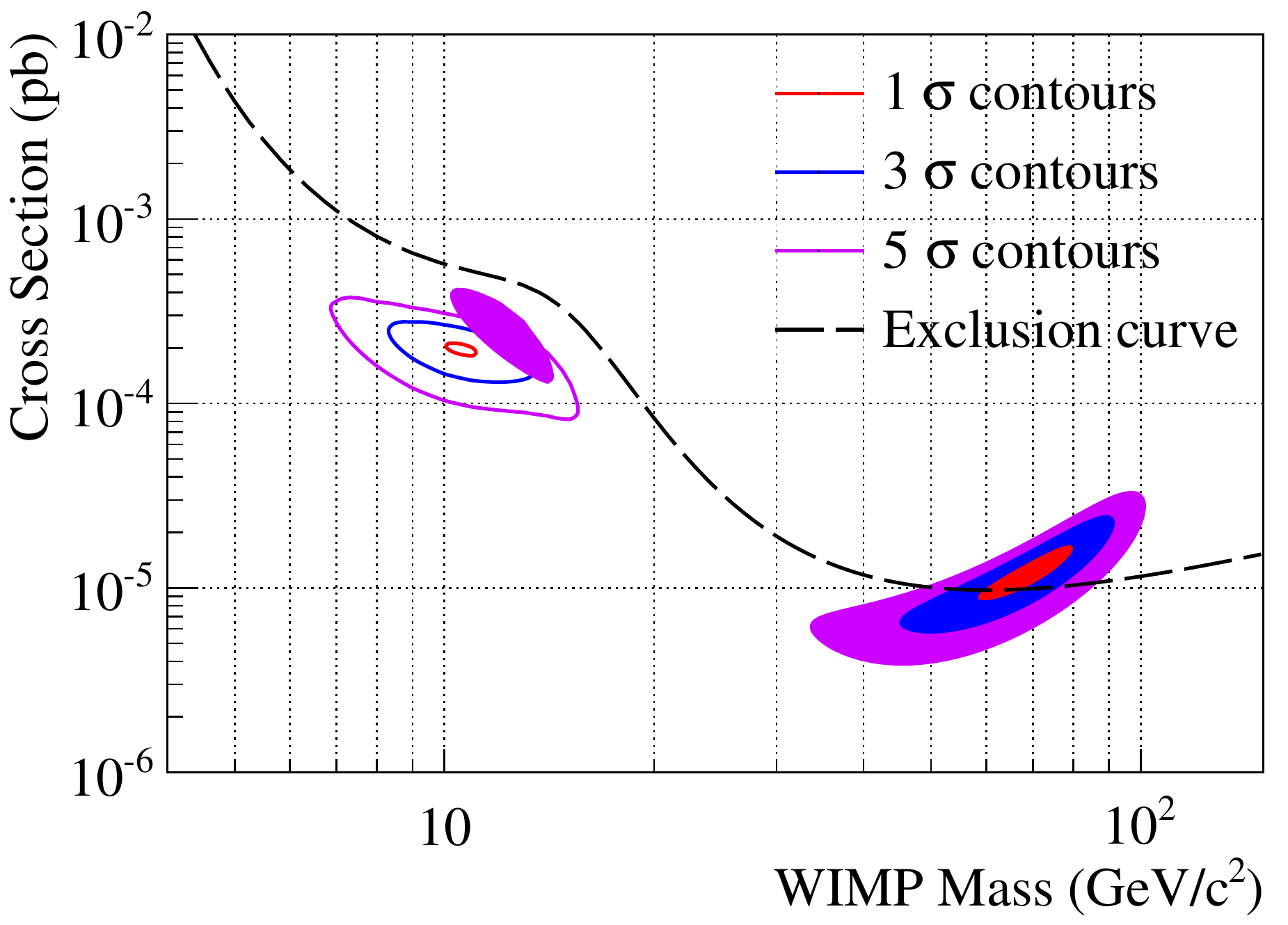}
\caption{The \dl\ 1\,$\sigma$, 3\,$\sigma$, 5\,$\sigma$ significance contours in the WIMP parameter space. The newly-obtained results using the quenching factors measured in this work are shown in color-filled regions, while the old results with the \dl\ quenching factors are shown in colored lines. The heavy-WIMP contours do not have significant change, but the 1\,$\sigma$ and 3\,$\sigma$ contours in the low-mass WIMP regions disappeared completely, disfavoring a light WIMP. The dashed line is the dark matter exclusion curve calculated using the overall \dl\ observed event rate~\cite{DAMA2008_FirstResults}. In the standard WIMP picture, WIMP parameters above this line would produce a nuclear recoil spectrum above the observed one in \dl\ at $\ge$\,1 energy bin.}
\label{fig:fitcontours}
\end{minipage}
\label{fig:damafit}
\end{figure}

Using a standard WIMP-halo model and considering only spin-independent WIMP-nucleon interactions~\cite{Lewin1996_DMMath, Savage2009_DAMAComp}, we evaluated how this new quenching measurement would impact the dark-matter interpretation of the \dl\ signal. Figure~\ref{fig:fit_spect} shows the WIMP fits of the \dl\ modulation data~\cite{DAMA2013_Phase1} with the new quenching factor values, compared with those using old quenching values. Since this work focuses on the Na quenching effects below 52\,\keVr (10.6\,\keVee), we restricted the fits to below 10\,\keVee\ energy, above which the \dl\ signal is consistent with zero modulation. As expected, the WIMP analysis yielded a global best fit with a heavy WIMP ($\sim$\,70\,GeV/$c^2$ mass) and a local best fit with a light WIMP ($\sim$\,10\,GeV/$c^2$ mass), as shown in Figure~\ref{fig:fitcontours}.  

With the new quenching factors, the light-WIMP fit (dominated by Na recoils) does not represent the data well, producing a $\chi^2_{\text{min}}$ of 36 with 14 degrees of freedom (P$<$\,0.01) in comparison with a $\chi^2_{\text{min}}$ of 19 (P$\sim$\,0.2) for the standard \dl\ fit with the old quenching value. Therefore, the low-mass WIMP region is strongly disfavored, as illustrated by the diminishing $\chi^2$ significance contours around 10\,GeV/$c^2$ in Figure~\ref{fig:fitcontours}. Moreover, by shifting the light-WIMP contours to larger WIMP-mass and cross-section values, the tension between \dl\ and other experiments increases in the standard WIMP picture. 

The high-mass-WIMP fit does not change significantly because the recoil signal is dominated by WIMP-iodine scatterings and the effect due to sodium quenching is negligible. Nonetheless, the best fit values in the high-mass WIMP region lead to a WIMP-interaction rate in \naitl\ higher than that observed in \dl\ around 2\,\keVee~\cite{DAMA2008_FirstResults}. The dark matter exclusion curve~\cite{Savage2009_DAMAComp} in Figure~\ref{fig:fitcontours} (dashed line) was calculated using the observed event rate in \dl\ between 2\,\keVee\ and 6\,\keVee, and in the standard WIMP picture, any WIMP parameters above this line would produce a nuclear recoil event rate higher than what was observed in \dl\ at one or more energy bins. More importantly, the iodine-dominated heavy-WIMP region has been excluded independently of WIMP models by the KIMS experiment using CsI(Tl) crystals~\cite{KIMS2012_CsILimits}. 

However, the standard WIMP models are known to have large uncertainties, and as discussed earlier, alternative dark matter theories may still be able to reconcile the experimental results~\cite{Feng2011_IVDM,Chiara2015_pseudoscalar}. A model-independent test of \dl, therefore, is best made with a \naitl\ experiment with lower background than that of \dl.

The lattice orientation of the \naitl\ crystal used in this experiment was not measured, so this measurement does not provide a sensitive test of the possible ion-channeling effect~\cite{DAMA2008_Channeling}. But, with the large number of Na-recoil angles measured and the fact that all quenching factors line up on a curve well below unit quenching (no quenching),  we do not observe any evidence for the channeling effect. Similarly, although the setup for this experiment was optimized to measure sodium recoils, iodine recoils should have been observed if the quenching factor were at the value measured by DAMA-LIBRA (0.09). Based on the absence of the iodine recoil peaks, we set an upper limit ($>$3\,$\sigma$) of 0.065 on the iodine recoil quenching factor at 10\,\keVr. 

%
%
%
%
   
\section{Conclusion}
\label{sec:conclusion}

We carried out an accurate measurement of the relative \naitl\ scintillation efficiency for Na recoils induced by a pulsed neutron beam, covering an energy window from 3\,\keVr\ to 52\,\keVr\ (or 0.65 - 10.6\,\keVee, covering the whole \dl\ modulation signal region of 2 - 6\,\keVee). By using double-TOF cuts and double-PSD cuts, we suppressed the coincidence background in the measurement with a high efficiency and obtained the most accurate results to date.

The Na recoil quenching factors are found to decrease significantly at low energies, which caused the \dl\ modulation signal to be less compatible with a light-WIMP explanation in the standard WIMP picture. Although alternative models may still be able to reconcile the \dl\ signal with other experimental results, a model-independent test using ultra-high purity \naitl\ crystal detector is necessary to confirm or refute the \dl\ dark matter claim.   

\begin{acknowledgments}

We acknowledge the hospitality of the University of Notre Dame in hosting this experiment and lending necessary electronics for us to fulfill the measurement. We thank Stephen Pordes from Fermilab for sharing neutron detectors and electronics with us. Radiation Monitoring Devices, Inc. (RMD) provided the \naitl\ crystal used in this experiment, and former Princeton technical specialist Allan Nelson built the detector enclosure parts; we thank them for their contributions. We are grateful to Ben Loer for developing the daqman data-acquisition and analysis softwares that were adapted for this measurement. The SABRE \naitl\ program has been supported by NSF Grants PHY-0957083, PHY-1103987, PHY-1242625. This measurement was supported by the NSF Grants PHY-1242625 and PHY-1419765. Francis Froborg is supported by the Swiss National Science Foundation. Henning O. Back is supported by the NSF Grant PHY-1242585. Thomas Alexander is partially supported by the NSF Grant PHY-1211308. 

\end{acknowledgments}

\bibliography{bibliography}

\begin{thebibliography}{34}%
\makeatletter
\providecommand \@ifxundefined [1]{%
 \@ifx{#1\undefined}
}%
\providecommand \@ifnum [1]{%
 \ifnum #1\expandafter \@firstoftwo
 \else \expandafter \@secondoftwo
 \fi
}%
\providecommand \@ifx [1]{%
 \ifx #1\expandafter \@firstoftwo
 \else \expandafter \@secondoftwo
 \fi
}%
\providecommand \natexlab [1]{#1}%
\providecommand \enquote  [1]{``#1''}%
\providecommand \bibnamefont  [1]{#1}%
\providecommand \bibfnamefont [1]{#1}%
\providecommand \citenamefont [1]{#1}%
\providecommand \href@noop [0]{\@secondoftwo}%
\providecommand \href [0]{\begingroup \@sanitize@url \@href}%
\providecommand \@href[1]{\@@startlink{#1}\@@href}%
\providecommand \@@href[1]{\endgroup#1\@@endlink}%
\providecommand \@sanitize@url [0]{\catcode `\\12\catcode `\$12\catcode
  `\&12\catcode `\#12\catcode `\^12\catcode `\_12\catcode `\%12\relax}%
\providecommand \@@startlink[1]{}%
\providecommand \@@endlink[0]{}%
\providecommand \url  [0]{\begingroup\@sanitize@url \@url }%
\providecommand \@url [1]{\endgroup\@href {#1}{\urlprefix }}%
\providecommand \urlprefix  [0]{URL }%
\providecommand \Eprint [0]{\href }%
\@ifxundefined \urlstyle {%
  \providecommand \doi  [0]{\begingroup \@sanitize@url \@doi}%
  \providecommand \@doi [1]{\endgroup \@@startlink {\doibase
  #1}doi:\discretionary {}{}{}#1\@@endlink }%
}{%
  \providecommand \doi  [0]{doi:\discretionary{}{}{}\begingroup
  \urlstyle{rm}\Url }%
}%
\providecommand \doibase [0]{http://dx.doi.org/}%
\providecommand \Doi [0]{\begingroup \@sanitize@url \@Doi }%
\providecommand \@Doi  [1]{\endgroup\@@startlink{\doibase#1}\@@Doi}%
\providecommand \@@Doi [1]{#1\@@endlink}%
\providecommand \selectlanguage [0]{\@gobble}%
\providecommand \bibinfo  [0]{\@secondoftwo}%
\providecommand \bibfield  [0]{\@secondoftwo}%
\providecommand \translation [1]{[#1]}%
\providecommand \BibitemOpen [0]{}%
\providecommand \bibitemStop [0]{}%
\providecommand \bibitemNoStop [0]{.\EOS\space}%
\providecommand \EOS [0]{\spacefactor3000\relax}%
\providecommand \BibitemShut  [1]{\csname bibitem#1\endcsname}%
\bibitem [{\citenamefont {Bernabei}\ \emph {et~al.}(2013)\citenamefont
  {Bernabei} \emph {et~al.}}]{DAMA2013_Phase1}%
  \BibitemOpen
  \bibfield  {author} {\bibinfo {author} {\bibfnamefont {R.}~\bibnamefont
  {Bernabei}} \emph {et~al.},\ }\Doi {10.1140/epjc/s10052-013-2648-7}
  {\bibfield  {journal} {\bibinfo  {journal} {The European Physical Journal
  C},\ }\textbf {\bibinfo {volume} {73}} (\bibinfo {year} {2013})},\ ISSN
  \bibinfo {issn} {1434-6044},\ \doi
  {10.1140/epjc/s10052-013-2648-7}\BibitemShut {NoStop}%
\bibitem [{\citenamefont {Fitzpatrick}\ \emph {et~al.}(2010)\citenamefont
  {Fitzpatrick}, \citenamefont {Hooper},\ and\ \citenamefont
  {Zurek}}]{Hooper2010_LightWIMP}%
  \BibitemOpen
  \bibfield  {author} {\bibinfo {author} {\bibfnamefont {A.~L.}\ \bibnamefont
  {Fitzpatrick}}, \bibinfo {author} {\bibfnamefont {D.}~\bibnamefont {Hooper}},
  \ and\ \bibinfo {author} {\bibfnamefont {K.~M.}\ \bibnamefont {Zurek}},\
  }\Doi {10.1103/PhysRevD.81.115005} {\bibfield  {journal} {\bibinfo  {journal}
  {Phys. Rev. D},\ }\textbf {\bibinfo {volume} {81}},\ \bibinfo {pages}
  {115005} (\bibinfo {year} {2010})}\BibitemShut {NoStop}%
\bibitem [{\citenamefont {Smith}\ and\ \citenamefont
  {Weiner}(2001)}]{Smith2001_InelasticDM}%
  \BibitemOpen
  \bibfield  {author} {\bibinfo {author} {\bibfnamefont {D.}~\bibnamefont
  {Smith}}\ and\ \bibinfo {author} {\bibfnamefont {N.}~\bibnamefont {Weiner}},\
  }\Doi {10.1103/PhysRevD.64.043502} {\bibfield  {journal} {\bibinfo  {journal}
  {Phys. Rev. D},\ }\textbf {\bibinfo {volume} {64}},\ \bibinfo {pages}
  {043502} (\bibinfo {year} {2001})}\BibitemShut {NoStop}%
\bibitem [{\citenamefont {Bernabei}\ \emph {et~al.}(2002)\citenamefont
  {Bernabei}, \citenamefont {Belli}, \citenamefont {Cerulli}, \citenamefont
  {Montecchia}, \citenamefont {Amato}, \citenamefont {Incicchitti},
  \citenamefont {Prosperi}, \citenamefont {Dai}, \citenamefont {He},
  \citenamefont {Kuang},\ and\ \citenamefont {Ma}}]{Bernabei2002_InelasticDM}%
  \BibitemOpen
  \bibfield  {author} {\bibinfo {author} {\bibfnamefont {R.}~\bibnamefont
  {Bernabei}}, \bibinfo {author} {\bibfnamefont {P.}~\bibnamefont {Belli}},
  \bibinfo {author} {\bibfnamefont {R.}~\bibnamefont {Cerulli}}, \bibinfo
  {author} {\bibfnamefont {F.}~\bibnamefont {Montecchia}}, \bibinfo {author}
  {\bibfnamefont {M.}~\bibnamefont {Amato}}, \bibinfo {author} {\bibfnamefont
  {A.}~\bibnamefont {Incicchitti}}, \bibinfo {author} {\bibfnamefont
  {D.}~\bibnamefont {Prosperi}}, \bibinfo {author} {\bibfnamefont
  {C.}~\bibnamefont {Dai}}, \bibinfo {author} {\bibfnamefont {H.}~\bibnamefont
  {He}}, \bibinfo {author} {\bibfnamefont {H.}~\bibnamefont {Kuang}}, \ and\
  \bibinfo {author} {\bibfnamefont {J.}~\bibnamefont {Ma}},\ }\Doi
  {10.1007/s100520100854} {\bibfield  {journal} {\bibinfo  {journal} {The
  European Physical Journal C - Particles and Fields},\ }\textbf {\bibinfo
  {volume} {23}},\ \bibinfo {pages} {61} (\bibinfo {year} {2002})},\ ISSN
  \bibinfo {issn} {1434-6044}\BibitemShut {NoStop}%
\bibitem [{\citenamefont {Angle}\ \emph {et~al.}(2011)\citenamefont {Angle},
  \citenamefont {Aprile} \emph {et~al.}}]{XENON2011_LightWIMP}%
  \BibitemOpen
  \bibfield  {author} {\bibinfo {author} {\bibfnamefont {J.}~\bibnamefont
  {Angle}}, \bibinfo {author} {\bibfnamefont {E.}~\bibnamefont {Aprile}},
  \emph {et~al.},\ }\Doi {10.1103/PhysRevLett.107.051301} {\bibfield  {journal}
  {\bibinfo  {journal} {Phys. Rev. Lett.},\ }\textbf {\bibinfo {volume}
  {107}},\ \bibinfo {pages} {051301} (\bibinfo {year} {2011})}\BibitemShut
  {NoStop}%
\bibitem [{\citenamefont {Aprile}\ \emph {et~al.}(2012)\citenamefont {Aprile}
  \emph {et~al.}}]{XENON2012_225days}%
  \BibitemOpen
  \bibfield  {author} {\bibinfo {author} {\bibfnamefont {E.}~\bibnamefont
  {Aprile}} \emph {et~al.},\ }\Doi {10.1103/PhysRevLett.109.181301} {\bibfield
  {journal} {\bibinfo  {journal} {Phys. Rev. Lett.},\ }\textbf {\bibinfo
  {volume} {109}},\ \bibinfo {pages} {181301} (\bibinfo {year}
  {2012})}\BibitemShut {NoStop}%
\bibitem [{\citenamefont {Agnese}\ \emph {et~al.}(2013)\citenamefont {Agnese}
  \emph {et~al.}}]{CDMS2013_SiDM}%
  \BibitemOpen
  \bibfield  {author} {\bibinfo {author} {\bibfnamefont {R.}~\bibnamefont
  {Agnese}} \emph {et~al.},\ }\Doi {10.1103/PhysRevLett.111.251301} {\bibfield
  {journal} {\bibinfo  {journal} {Phys. Rev. Lett.},\ }\textbf {\bibinfo
  {volume} {111}},\ \bibinfo {pages} {251301} (\bibinfo {year}
  {2013})}\BibitemShut {NoStop}%
\bibitem [{\citenamefont {{\relax LUX
  Collaboration}}(2014)}]{LUX2014_DMResult}%
  \BibitemOpen
  \bibfield  {author} {\bibinfo {author} {\bibnamefont {{\relax LUX
  Collaboration}}},\ }\Doi {10.1103/PhysRevLett.112.091303} {\bibfield
  {journal} {\bibinfo  {journal} {Phys. Rev. Lett.},\ }\textbf {\bibinfo
  {volume} {112}},\ \bibinfo {pages} {091303} (\bibinfo {year}
  {2014})}\BibitemShut {NoStop}%
\bibitem [{\citenamefont {Feng}\ \emph {et~al.}(2011)\citenamefont {Feng},
  \citenamefont {Kumar}, \citenamefont {Marfatia},\ and\ \citenamefont
  {Sanford}}]{Feng2011_IVDM}%
  \BibitemOpen
  \bibfield  {author} {\bibinfo {author} {\bibfnamefont {J.~L.}\ \bibnamefont
  {Feng}}, \bibinfo {author} {\bibfnamefont {J.}~\bibnamefont {Kumar}},
  \bibinfo {author} {\bibfnamefont {D.}~\bibnamefont {Marfatia}}, \ and\
  \bibinfo {author} {\bibfnamefont {D.}~\bibnamefont {Sanford}},\ }\Doi
  {http://dx.doi.org/10.1016/j.physletb.2011.07.083} {\bibfield  {journal}
  {\bibinfo  {journal} {Physics Letters B},\ }\textbf {\bibinfo {volume}
  {703}},\ \bibinfo {pages} {124 } (\bibinfo {year} {2011})},\ ISSN \bibinfo
  {issn} {0370-2693}\BibitemShut {NoStop}%
\bibitem [{\citenamefont {Arina}\ \emph {et~al.}(2015)\citenamefont {Arina},
  \citenamefont {Del~Nobile},\ and\ \citenamefont
  {Panci}}]{Chiara2015_pseudoscalar}%
  \BibitemOpen
  \bibfield  {author} {\bibinfo {author} {\bibfnamefont {C.}~\bibnamefont
  {Arina}}, \bibinfo {author} {\bibfnamefont {E.}~\bibnamefont {Del~Nobile}}, \
  and\ \bibinfo {author} {\bibfnamefont {P.}~\bibnamefont {Panci}},\ }\Doi
  {10.1103/PhysRevLett.114.011301} {\bibfield  {journal} {\bibinfo  {journal}
  {Phys. Rev. Lett.},\ }\textbf {\bibinfo {volume} {114}},\ \bibinfo {pages}
  {011301} (\bibinfo {year} {2015})}\BibitemShut {NoStop}%
\bibitem [{\citenamefont {Bernabei}\ \emph {et~al.}(1996)\citenamefont
  {Bernabei}, \citenamefont {Belli}, \citenamefont {Landoni}, \citenamefont
  {Montecchia}, \citenamefont {Di}, \citenamefont {Incicchitti}, \citenamefont
  {Prosperi}, \citenamefont {Bacci}, \citenamefont {C.J.}, \citenamefont
  {L.K.}, \citenamefont {Kuang}, \citenamefont {Ma}, \citenamefont {Angelone},
  \citenamefont {Bastistoni},\ and\ \citenamefont {Pillon}}]{DamaQuench1996}%
  \BibitemOpen
  \bibfield  {author} {\bibinfo {author} {\bibfnamefont {R.}~\bibnamefont
  {Bernabei}}, \bibinfo {author} {\bibfnamefont {P.}~\bibnamefont {Belli}},
  \bibinfo {author} {\bibfnamefont {V.}~\bibnamefont {Landoni}}, \bibinfo
  {author} {\bibfnamefont {F.}~\bibnamefont {Montecchia}}, \bibinfo {author}
  {\bibfnamefont {N.~W.}\ \bibnamefont {Di}}, \bibinfo {author} {\bibfnamefont
  {A.}~\bibnamefont {Incicchitti}}, \bibinfo {author} {\bibfnamefont
  {D.}~\bibnamefont {Prosperi}}, \bibinfo {author} {\bibfnamefont
  {C.}~\bibnamefont {Bacci}}, \bibinfo {author} {\bibfnamefont
  {D.}~\bibnamefont {C.J.}}, \bibinfo {author} {\bibfnamefont {D.}~\bibnamefont
  {L.K.}}, \bibinfo {author} {\bibfnamefont {H.}~\bibnamefont {Kuang}},
  \bibinfo {author} {\bibfnamefont {J.}~\bibnamefont {Ma}}, \bibinfo {author}
  {\bibfnamefont {M.}~\bibnamefont {Angelone}}, \bibinfo {author}
  {\bibfnamefont {P.}~\bibnamefont {Bastistoni}}, \ and\ \bibinfo {author}
  {\bibfnamefont {M.}~\bibnamefont {Pillon}},\ }\Doi
  {http://dx.doi.org/10.1016/S0370-2693(96)80020-7} {\bibfield  {journal}
  {\bibinfo  {journal} {Physics Letters B},\ }\textbf {\bibinfo {volume}
  {389}},\ \bibinfo {pages} {757 } (\bibinfo {year} {1996})},\ ISSN \bibinfo
  {issn} {0370-2693}\BibitemShut {NoStop}%
\bibitem [{\citenamefont {Spooner}\ \emph {et~al.}(1994)\citenamefont
  {Spooner}, \citenamefont {Davies}, \citenamefont {Davies}, \citenamefont
  {Pyle}, \citenamefont {Bucknell}, \citenamefont {Squier}, \citenamefont
  {Lewin},\ and\ \citenamefont {Smith}}]{Spooner1994}%
  \BibitemOpen
  \bibfield  {author} {\bibinfo {author} {\bibfnamefont {N.}~\bibnamefont
  {Spooner}}, \bibinfo {author} {\bibfnamefont {G.}~\bibnamefont {Davies}},
  \bibinfo {author} {\bibfnamefont {J.}~\bibnamefont {Davies}}, \bibinfo
  {author} {\bibfnamefont {G.}~\bibnamefont {Pyle}}, \bibinfo {author}
  {\bibfnamefont {T.}~\bibnamefont {Bucknell}}, \bibinfo {author}
  {\bibfnamefont {G.}~\bibnamefont {Squier}}, \bibinfo {author} {\bibfnamefont
  {J.}~\bibnamefont {Lewin}}, \ and\ \bibinfo {author} {\bibfnamefont
  {P.}~\bibnamefont {Smith}},\ }\Doi
  {http://dx.doi.org/10.1016/0370-2693(94)90343-3} {\bibfield  {journal}
  {\bibinfo  {journal} {Physics Letters B},\ }\textbf {\bibinfo {volume}
  {321}},\ \bibinfo {pages} {156 } (\bibinfo {year} {1994})},\ ISSN \bibinfo
  {issn} {0370-2693}\BibitemShut {NoStop}%
\bibitem [{\citenamefont {Tovey}\ \emph {et~al.}(1998)\citenamefont {Tovey},
  \citenamefont {Kudryavtsev}, \citenamefont {Lehner}, \citenamefont
  {McMillan}, \citenamefont {Peak}, \citenamefont {Roberts}, \citenamefont
  {Spooner},\ and\ \citenamefont {Lewin}}]{Tovey1998}%
  \BibitemOpen
  \bibfield  {author} {\bibinfo {author} {\bibfnamefont {D.}~\bibnamefont
  {Tovey}}, \bibinfo {author} {\bibfnamefont {V.}~\bibnamefont {Kudryavtsev}},
  \bibinfo {author} {\bibfnamefont {M.}~\bibnamefont {Lehner}}, \bibinfo
  {author} {\bibfnamefont {J.}~\bibnamefont {McMillan}}, \bibinfo {author}
  {\bibfnamefont {C.}~\bibnamefont {Peak}}, \bibinfo {author} {\bibfnamefont
  {J.}~\bibnamefont {Roberts}}, \bibinfo {author} {\bibfnamefont
  {N.}~\bibnamefont {Spooner}}, \ and\ \bibinfo {author} {\bibfnamefont
  {J.}~\bibnamefont {Lewin}},\ }\Doi
  {http://dx.doi.org/10.1016/S0370-2693(98)00643-1} {\bibfield  {journal}
  {\bibinfo  {journal} {Physics Letters B},\ }\textbf {\bibinfo {volume}
  {433}},\ \bibinfo {pages} {150 } (\bibinfo {year} {1998})},\ ISSN \bibinfo
  {issn} {0370-2693}\BibitemShut {NoStop}%
\bibitem [{\citenamefont {Gerbier}\ \emph {et~al.}(1999)\citenamefont
  {Gerbier}, \citenamefont {Mallet}, \citenamefont {Mosca}, \citenamefont
  {Tao}, \citenamefont {Chambon}, \citenamefont {Chazal}, \citenamefont
  {J{\'e}sus}, \citenamefont {Drain}, \citenamefont {Messous},\ and\
  \citenamefont {Pastor}}]{Gerbier1999}%
  \BibitemOpen
  \bibfield  {author} {\bibinfo {author} {\bibfnamefont {G.}~\bibnamefont
  {Gerbier}}, \bibinfo {author} {\bibfnamefont {J.}~\bibnamefont {Mallet}},
  \bibinfo {author} {\bibfnamefont {L.}~\bibnamefont {Mosca}}, \bibinfo
  {author} {\bibfnamefont {C.}~\bibnamefont {Tao}}, \bibinfo {author}
  {\bibfnamefont {B.}~\bibnamefont {Chambon}}, \bibinfo {author} {\bibfnamefont
  {V.}~\bibnamefont {Chazal}}, \bibinfo {author} {\bibfnamefont {M.~D.}\
  \bibnamefont {J{\'e}sus}}, \bibinfo {author} {\bibfnamefont {D.}~\bibnamefont
  {Drain}}, \bibinfo {author} {\bibfnamefont {Y.}~\bibnamefont {Messous}}, \
  and\ \bibinfo {author} {\bibfnamefont {C.}~\bibnamefont {Pastor}},\ }\Doi
  {http://dx.doi.org/10.1016/S0927-6505(99)00004-3} {\bibfield  {journal}
  {\bibinfo  {journal} {Astroparticle Physics},\ }\textbf {\bibinfo {volume}
  {11}},\ \bibinfo {pages} {287 } (\bibinfo {year} {1999})},\ ISSN \bibinfo
  {issn} {0927-6505}\BibitemShut {NoStop}%
\bibitem [{\citenamefont {Jagemann}\ \emph {et~al.}(2006)\citenamefont
  {Jagemann}, \citenamefont {Feilitzsch},\ and\ \citenamefont
  {Jochum}}]{Jagemann2006}%
  \BibitemOpen
  \bibfield  {author} {\bibinfo {author} {\bibfnamefont {T.}~\bibnamefont
  {Jagemann}}, \bibinfo {author} {\bibfnamefont {F.}~\bibnamefont
  {Feilitzsch}}, \ and\ \bibinfo {author} {\bibfnamefont {J.}~\bibnamefont
  {Jochum}},\ }\Doi {http://dx.doi.org/10.1016/j.nima.2006.03.029} {\bibfield
  {journal} {\bibinfo  {journal} {Nuclear Instruments and Methods in Physics
  Research Section A: Accelerators, Spectrometers, Detectors and Associated
  Equipment},\ }\textbf {\bibinfo {volume} {564}},\ \bibinfo {pages} {549 }
  (\bibinfo {year} {2006})},\ ISSN \bibinfo {issn} {0168-9002}\BibitemShut
  {NoStop}%
\bibitem [{\citenamefont {Simon}\ \emph {et~al.}(2003)\citenamefont {Simon}
  \emph {et~al.}}]{Simon2003}%
  \BibitemOpen
  \bibfield  {author} {\bibinfo {author} {\bibfnamefont {E.}~\bibnamefont
  {Simon}} \emph {et~al.},\ }\Doi
  {http://dx.doi.org/10.1016/S0168-9002(03)01438-4} {\bibfield  {journal}
  {\bibinfo  {journal} {Nuclear Instruments and Methods in Physics Research
  Section A: Accelerators, Spectrometers, Detectors and Associated Equipment},\
  }\textbf {\bibinfo {volume} {507}},\ \bibinfo {pages} {643 } (\bibinfo {year}
  {2003})},\ ISSN \bibinfo {issn} {0168-9002}\BibitemShut {NoStop}%
\bibitem [{\citenamefont {Chagani}\ \emph {et~al.}(2008)\citenamefont
  {Chagani}, \citenamefont {Majewski}, \citenamefont {Daw}, \citenamefont
  {Kudryavtsev},\ and\ \citenamefont {Spooner}}]{Chagani2008}%
  \BibitemOpen
  \bibfield  {author} {\bibinfo {author} {\bibfnamefont {H.}~\bibnamefont
  {Chagani}}, \bibinfo {author} {\bibfnamefont {P.}~\bibnamefont {Majewski}},
  \bibinfo {author} {\bibfnamefont {E.~J.}\ \bibnamefont {Daw}}, \bibinfo
  {author} {\bibfnamefont {V.~A.}\ \bibnamefont {Kudryavtsev}}, \ and\ \bibinfo
  {author} {\bibfnamefont {N.~J.~C.}\ \bibnamefont {Spooner}},\ }\href
  {http://stacks.iop.org/1748-0221/3/i=06/a=P06003} {\bibfield  {journal}
  {\bibinfo  {journal} {Journal of Instrumentation},\ }\textbf {\bibinfo
  {volume} {3}},\ \bibinfo {pages} {P06003} (\bibinfo {year}
  {2008})}\BibitemShut {NoStop}%
\bibitem [{\citenamefont {Collar}(2013)}]{Collar2013}%
  \BibitemOpen
  \bibfield  {author} {\bibinfo {author} {\bibfnamefont {J.~I.}\ \bibnamefont
  {Collar}},\ }\Doi {10.1103/PhysRevC.88.035806} {\bibfield  {journal}
  {\bibinfo  {journal} {Phys. Rev. C},\ }\textbf {\bibinfo {volume} {88}},\
  \bibinfo {pages} {035806} (\bibinfo {year} {2013})}\BibitemShut {NoStop}%
\bibitem [{\citenamefont {Kim}\ \emph {et~al.}(2015)\citenamefont {Kim} \emph
  {et~al.}}]{Kim2015_NaI}%
  \BibitemOpen
  \bibfield  {author} {\bibinfo {author} {\bibfnamefont {K.}~\bibnamefont
  {Kim}} \emph {et~al.},\ }\Doi
  {http://dx.doi.org/10.1016/j.astropartphys.2014.10.004} {\bibfield  {journal}
  {\bibinfo  {journal} {Astroparticle Physics},\ }\textbf {\bibinfo {volume}
  {62}},\ \bibinfo {pages} {249 } (\bibinfo {year} {2015})},\ ISSN \bibinfo
  {issn} {0927-6505}\BibitemShut {NoStop}%
\bibitem [{\citenamefont {Cherwinka}\ and\ \citenamefont
  {others.}(2014)}]{DMIce2014_firstdata}%
  \BibitemOpen
  \bibfield  {author} {\bibinfo {author} {\bibfnamefont {J.}~\bibnamefont
  {Cherwinka}}\ and\ \bibinfo {author} {\bibnamefont {others.}} (\bibinfo
  {collaboration} {(DM\char21{}Ice Collaboration)}),\ }\Doi
  {10.1103/PhysRevD.90.092005} {\bibfield  {journal} {\bibinfo  {journal}
  {Phys. Rev. D},\ }\textbf {\bibinfo {volume} {90}},\ \bibinfo {pages}
  {092005} (\bibinfo {year} {2014})}\BibitemShut {NoStop}%
\bibitem [{\citenamefont {Amare}\ \emph {et~al.}(2012)\citenamefont {Amare}
  \emph {et~al.}}]{ANAIS2012_ANAIS0}%
  \BibitemOpen
  \bibfield  {author} {\bibinfo {author} {\bibfnamefont {J.}~\bibnamefont
  {Amare}} \emph {et~al.},\ }\Doi
  {http://dx.doi.org/10.1088/1742-6596/375/1/012026} {\bibfield  {journal}
  {\bibinfo  {journal} {Journal of Physics: Conference Series},\ }\textbf
  {\bibinfo {volume} {375}},\ \bibinfo {pages} {012026} (\bibinfo {year}
  {2012})}\BibitemShut {NoStop}%
\bibitem [{\citenamefont {Alner}\ \emph {et~al.}(2005)\citenamefont {Alner}
  \emph {et~al.}}]{NaIAD2005_WIMP}%
  \BibitemOpen
  \bibfield  {author} {\bibinfo {author} {\bibfnamefont {G.}~\bibnamefont
  {Alner}} \emph {et~al.},\ }\Doi
  {http://dx.doi.org/10.1016/j.physletb.2000.09.001} {\bibfield  {journal}
  {\bibinfo  {journal} {Physics Letters B},\ }\textbf {\bibinfo {volume}
  {616}},\ \bibinfo {pages} {17 } (\bibinfo {year} {2005})},\ ISSN \bibinfo
  {issn} {0370-2693}\BibitemShut {NoStop}%
\bibitem [{SAB(2013)}]{SABRE2013_TAUP}%
  \BibitemOpen
  \href@noop {} {\emph {\bibinfo {title} {SABRE: A new NaI(Tl) dark matter
  direct detection experiment}}}\ (\bibinfo {year} {2013})\ \bibinfo {note}
  {{TAUP}2013 proceeding, to be published}\BibitemShut {NoStop}%
\bibitem [{\citenamefont {{\relax The SCENE
  Collaboration}}(2013)}]{SCENE2013_FieldQuenching}%
  \BibitemOpen
  \bibfield  {author} {\bibinfo {author} {\bibnamefont {{\relax The SCENE
  Collaboration}}},\ }\Doi {10.1103/PhysRevD.88.092006} {\bibfield  {journal}
  {\bibinfo  {journal} {Phys. Rev. D},\ }\textbf {\bibinfo {volume} {88}},\
  \bibinfo {pages} {092006} (\bibinfo {year} {2013})}\BibitemShut {NoStop}%
\bibitem [{\citenamefont {{\relax The SCENE
  Collaboration}}(2014)}]{SCENE2014_Quenching}%
  \BibitemOpen
  \bibfield  {author} {\bibinfo {author} {\bibnamefont {{\relax The SCENE
  Collaboration}}},\ }\href {http://arxiv.org/abs/1406.4825} {\bibfield
  {journal} {\bibinfo  {journal} {{arXiv:1406.4825}}} (\bibinfo {year}
  {2014})}\BibitemShut {NoStop}%
\bibitem [{\citenamefont {Burke}\ \emph {et~al.}(1974)\citenamefont {Burke},
  \citenamefont {Lunnon},\ and\ \citenamefont {Lefevre}}]{Burke1974_LipnBe}%
  \BibitemOpen
  \bibfield  {author} {\bibinfo {author} {\bibfnamefont {C.~A.}\ \bibnamefont
  {Burke}}, \bibinfo {author} {\bibfnamefont {M.~T.}\ \bibnamefont {Lunnon}}, \
  and\ \bibinfo {author} {\bibfnamefont {H.~W.}\ \bibnamefont {Lefevre}},\
  }\Doi {10.1103/PhysRevC.10.1299} {\bibfield  {journal} {\bibinfo  {journal}
  {Phys. Rev. C},\ }\textbf {\bibinfo {volume} {10}},\ \bibinfo {pages} {1299}
  (\bibinfo {year} {1974})}\BibitemShut {NoStop}%
\bibitem [{\citenamefont {Knoll}(2010)}]{knoll_raddetection}%
  \BibitemOpen
  \bibfield  {author} {\bibinfo {author} {\bibfnamefont {G.~F.}\ \bibnamefont
  {Knoll}},\ }\href@noop {} {\emph {\bibinfo {title} {{Radiation detection and
  measurement; 4th ed.}}}}\ (\bibinfo  {publisher} {Wiley},\ \bibinfo {address}
  {New York, NY},\ \bibinfo {year} {2010})\BibitemShut {NoStop}%
\bibitem [{\citenamefont {Bernabei}\ \emph
  {et~al.}(2008){\natexlab{a}}\citenamefont {Bernabei} \emph
  {et~al.}}]{DAMA2008_apparatus}%
  \BibitemOpen
  \bibfield  {author} {\bibinfo {author} {\bibfnamefont {R.}~\bibnamefont
  {Bernabei}} \emph {et~al.},\ }\href {http://arxiv.org/abs/0804.2738v1}
  {\bibfield  {journal} {\bibinfo  {journal} {{arXiv:0804.2738v1}}} (\bibinfo
  {year} {2008}{\natexlab{a}})}\BibitemShut {NoStop}%
\bibitem [{\citenamefont {Agostinelli}\ \emph {et~al.}(2003)\citenamefont
  {Agostinelli} \emph {et~al.}}]{geant4}%
  \BibitemOpen
  \bibfield  {author} {\bibinfo {author} {\bibfnamefont {S.}~\bibnamefont
  {Agostinelli}} \emph {et~al.},\ }\href@noop {} {\bibfield  {journal}
  {\bibinfo  {journal} {Nucl. Instrum. Meth. A},\ }\textbf {\bibinfo {volume}
  {506}},\ \bibinfo {pages} {250 } (\bibinfo {year} {2003})}\BibitemShut
  {NoStop}%
\bibitem [{\citenamefont {Lewin}\ and\ \citenamefont
  {Smith}(1996)}]{Lewin1996_DMMath}%
  \BibitemOpen
  \bibfield  {author} {\bibinfo {author} {\bibfnamefont {J.~D.}\ \bibnamefont
  {Lewin}}\ and\ \bibinfo {author} {\bibfnamefont {P.~F.}\ \bibnamefont
  {Smith}},\ }\Doi {10.1016/S0927-6505(96)00047-3} {\bibfield  {journal}
  {\bibinfo  {journal} {Astropart. Phys.},\ }\textbf {\bibinfo {volume} {6}},\
  \bibinfo {pages} {87 } (\bibinfo {year} {1996})},\ ISSN \bibinfo {issn}
  {0927-6505}\BibitemShut {NoStop}%
\bibitem [{\citenamefont {Savage}\ \emph {et~al.}(2009)\citenamefont {Savage},
  \citenamefont {Gelmini}, \citenamefont {Gondolo},\ and\ \citenamefont
  {Freese}}]{Savage2009_DAMAComp}%
  \BibitemOpen
  \bibfield  {author} {\bibinfo {author} {\bibfnamefont {C.}~\bibnamefont
  {Savage}}, \bibinfo {author} {\bibfnamefont {G.}~\bibnamefont {Gelmini}},
  \bibinfo {author} {\bibfnamefont {P.}~\bibnamefont {Gondolo}}, \ and\
  \bibinfo {author} {\bibfnamefont {K.}~\bibnamefont {Freese}},\ }\href
  {http://stacks.iop.org/1475-7516/2009/i=04/a=010} {\bibfield  {journal}
  {\bibinfo  {journal} {Journal of Cosmology and Astroparticle Physics},\
  }\textbf {\bibinfo {volume} {2009}},\ \bibinfo {pages} {010} (\bibinfo {year}
  {2009})}\BibitemShut {NoStop}%
\bibitem [{\citenamefont {Bernabei}\ \emph
  {et~al.}(2008){\natexlab{b}}\citenamefont {Bernabei} \emph
  {et~al.}}]{DAMA2008_FirstResults}%
  \BibitemOpen
  \bibfield  {author} {\bibinfo {author} {\bibfnamefont {R.}~\bibnamefont
  {Bernabei}} \emph {et~al.},\ }\Doi {10.1140/epjc/s10052-008-0662-y}
  {\bibfield  {journal} {\bibinfo  {journal} {The European Physical Journal
  C},\ }\textbf {\bibinfo {volume} {56}},\ \bibinfo {pages} {333} (\bibinfo
  {year} {2008}{\natexlab{b}})},\ ISSN \bibinfo {issn} {1434-6044}\BibitemShut
  {NoStop}%
\bibitem [{\citenamefont {Kim}\ \emph {et~al.}(2012)\citenamefont {Kim} \emph
  {et~al.}}]{KIMS2012_CsILimits}%
  \BibitemOpen
  \bibfield  {author} {\bibinfo {author} {\bibfnamefont {S.~C.}\ \bibnamefont
  {Kim}} \emph {et~al.},\ }\Doi {10.1103/PhysRevLett.108.181301} {\bibfield
  {journal} {\bibinfo  {journal} {Phys. Rev. Lett.},\ }\textbf {\bibinfo
  {volume} {108}},\ \bibinfo {pages} {181301} (\bibinfo {year}
  {2012})}\BibitemShut {NoStop}%
\bibitem [{\citenamefont {Bernabei}\ \emph
  {et~al.}(2008){\natexlab{c}}\citenamefont {Bernabei} \emph
  {et~al.}}]{DAMA2008_Channeling}%
  \BibitemOpen
  \bibfield  {author} {\bibinfo {author} {\bibfnamefont {R.}~\bibnamefont
  {Bernabei}} \emph {et~al.},\ }\Doi {10.1140/epjc/s10052-007-0479-0}
  {\bibfield  {journal} {\bibinfo  {journal} {The European Physical Journal
  C},\ }\textbf {\bibinfo {volume} {53}},\ \bibinfo {pages} {205} (\bibinfo
  {year} {2008}{\natexlab{c}})},\ ISSN \bibinfo {issn} {1434-6044}\BibitemShut
  {NoStop}%
\end{thebibliography}%

\end{document}